\documentclass[review]{elsarticle}
\usepackage{lineno}
\usepackage{geometry}
\usepackage{booktabs}
\usepackage{threeparttable}
\usepackage{amsmath}
\usepackage{amsthm}
\usepackage{mathrsfs}
\usepackage{amssymb}
\usepackage{graphics}
\usepackage{graphicx}
\usepackage{multirow}
\usepackage{subfig}
\usepackage{bm}
\usepackage{algorithm}
\usepackage{array}
\usepackage{algpseudocode}
\usepackage{graphics}
\usepackage[colorlinks,linkcolor=red, anchorcolor=blue,citecolor=blue]{hyperref}
\usepackage{lscape}
\usepackage{pdflscape}

\modulolinenumbers[1]
\graphicspath{{./figs/}}
\biboptions{authoryear}
\DeclareMathOperator*{\argmax}{arg\,max}
\begin{document}
% \linenumbers
 \begin{frontmatter}
\title{Go left or right? Explore the side preference behavior with \\ circle antipode experiments}
% \title{Investigation of Pedestrian Dynamics in Circle Antipode Experiments: Side Preference} 

\author[1]{Yao Xiao}
\author[2]{Ziyou Gao*}
\ead{zygao@bjtu.edu.cn}
\author[2]{Rui Jiang*}
\ead{jiangrui@bjtu.edu.cn}
\author[2]{Qinxia Huang}
\author[1]{Hai Yang}
\address[1]{Department of Civil and Environmental Engineering, The Hong Kong University of Science and Technology, Clear Water Bay, Kowloon, Hong Kong, China}
\address[2]{School of traffic and transportation, Beijing Jiaotong University, Beijing, China}

% \cortext[mycorrespondingauthor]{E-mail address: xiaoy@bjtu.edu.cn} 

% \fntext[myfootnote]{Since 1880.}
% %% or include affiliations in footnotes:
% \author[mymainaddress,mysecondaryaddress]{Elsevier Inc}
%\author[1222]{\corref{mycorrespondingauthor}}
% \address[mysecondaryaddress]{360 Park Avenue South, New York}

\begin{abstract}
Side preference is a critical behavior in conflict handling, and here the behavior is investigated with pedestrian trajectories in circle antipode experiments that own both conflicting and symmetrical participant situations. In the series of experiments, more participants(around 70\%) prefer to walk on the right side, and the statistical analyses reveal that factors such as handedness, gender, and height have no significant impacts. Further investigations show that most pedestrians actually make the side choices at the very beginning, and empirical results suggest that selecting the dominate side preference (right side in our experiments) can benefit the individual movement efficiency. To reflect the realistic side preference characteristics in simulations, a Voronoi diagram based model is introduced as well as a side preference parameter with normal distribution is formulated and calibrated. Further simulations prove that the modified model is able to reproduce realistic side preference behaviors in circle antipode experiments and other common situations. 
\end{abstract}

% \begin{keyword}
% Pedestrian Dynamics, Circle Antipode Experiment, Pedestrian Trajectories, Model Evaluation
% %\texttt{elsarticle.cls}\sep \LaTeX\sep Elsevier \sep template
% %\MSC[2010] 00-01\sep 99-00
% \end{keyword}
 \end{frontmatter}

\section{Introduction}\label{section 1}

With the development of societies, the pedestrian researches have drawn great attentions from publics and researchers in recent decades. Nowadays, especially during the peak-hours in daily work or in large-scale activities, lots of problems such as safety issues, efficiency issues and comfort issues frequently emerge. The study of pedestrian dynamics can help to guide the design of pedestrian facilities in public places and the organization of crowds in activities.

The experiment with a controlled environment and human subjects is one of the most widely used methods in pedestrian researches. Current controlled experiments mainly include the experiments of the practical common scenes such as corridor and bottleneck. The single-file experiment is a most fundamental scene in the corridor experiments. \cite{Seyfried2005} investigated the single file movement of pedestrians and discovered the linear relationship between velocity and the reciprocal of density. Similar rules were found by \cite{Jelic2012a,Jelic2012}, and further investigations about the microscopic step behaviors shown the negative correlation between safety distance and step time variance. The density-velocity-flow relationship is one of the most popular experimental results in corridor experiments \citep{Weidmann1993,Helbing2007,Zhang2011}. The congestion reaction behaviors are concerned in bottleneck experiments, and the capacity is the most widely-used experiment index. \cite{Nicolas2017} further investigated the influence of pedestrian heterogeneity and found that the increasing ratios of aggressive pedestrians could raise the capacity and bring fluctuation to flow rate. \cite{Kruechten2017} investigated the crowd behaviors and found that the growth crowd scale was beneficial for the evacuation. Other frequently-used experiments include multi-directional experiments (bi-directional experiments \citep{Weidmann1993,Lam2002,Guo2012,Zhang2012,Cao2017} and Y/X shape intersect\citep{Cao2017,Daamen2003,Helbing2005,Asano2010,Shi2016}), corner experiments\citep{Dias2014}, and route choice experiments\citep{Guo2012,Haghani2017,Wagoum2017,Haghani2019}. In these traditional experiments, the quantitative analysis is usually limited due to the situation differences among participants, therefore some behaviors are not so easy to be investigated. In the circle antipode experiments\citep{Xiao2019}, the participants are uniformly distributed on the circle at the initial stage, and they are simultaneously required to leave for the antipodal position as fast as possible. Conflicts frequently occur in the central region during the experiments, and the participants own symmetric experiment situations. These two characteristics lay the base for the investigations of pedestrian behaviors and further self-organized phenomenon.

The self-organization phenomenon is a macroscopic orderly phenomenon, and it spontaneously formed by pedestrian crowds in the process of movement. Widely-used phenomena include lane formation\citep{Seyfried2009,Helbing2005,Helbing2009}, faster-is-slower effect \citep{Helbing2009}, zipper effect \citep{Hoogendoorn2005}, stripe formation \citep{Helbing2009} , stop-and-go wave \citep{Helbing2009, Helbing2007} and so on. Among the phenomena, side preference \citep{Helbing2009} is about the preferred side choice (left or right side) when a pedestrian needs to deal with the conflict in the forward direction. It is a common self-organized phenomenon in the pedestrian crowd which occurs at the bi-directional flow situation, the overtaking situation, the obstacle situation and any other pedestrian situations with conflicts. The investigations of the side preference behavior mechanisms can contribute to the reproduction of practical crowd motion during simulation, in which the behavior is usually not considered but actually plays a critical role. Besides, more understandings about the behavior can even benefit the organization of crowds and the setting of public facilities in different aspects. Through a set of well-controlled experiments with simple avoidance requirements, the determination laws ruling their behaviors were investigated by \cite{Moussaid2009}. In the experiments, the side preference behavior is regarded as a kind of cultural bias, which indicates that the behaviors vary obviously in different regions. \cite{Jung2013} investigated the stereotypes of Koreans preferred walking direction with both observation and survey methods, and the results showed more pedestrians prefer the right side and the natural preferred direction should be a crucial determinant in traffic regulations.

To make a further quantitative investigation on the phenomenon in the work, a series of circle antipode experiments is applied for the side preference research. During the experiments, every pedestrian has to make the side preference decision since there is a crowded area in the shortest path. Due to the symmetric characteristics for participants, i.e. symmetric starting points, symmetric destination points, and symmetric situations, the experiment is also a favorable environment for analyzing side preference behaviors. To be specific, the side preference behaviors are easy to be quantitatively investigated in the experiments. Furthermore, a modified Voronoi model is proposed to reproduce the side preference behaviors in the circle antipode experiments, and a side preference parameter is formulated to simulate the pedestrian movements.

The rest of the paper is organized as follows. Section 2 mainly introduces the present of the circle antipode experiments. In Section 3, the side preference behaviors in the experiments are quantitatively investigated from the experiment level and individual level. In Section 4, a modified Voronoi model is introduced to reproduce the pedestrian behaviors in the circle antipode experiments and corridor situations. Section 5 is about the conclusions and discussions.

\section{Circle antipode experiment}\label{section 2}

The circle antipode experiment is a featured experiment to explore the pedestrian motion navigation and conflict avoidance behaviors. In the experiments, pedestrians were uniformly distributed on the circle at the very beginning, and they were required to leave for the antipode position of the circle at the same moment. Two remarkable characteristics are found in the experiments. First, the shortest routes intersect at the center point of the circle, where complicated conflicts frequently occur during the experiments. As a result, the conflict avoidance behaviors among pedestrians can be fully investigated. Second, the pedestrians in the experiments are symmetrical, i.e., symmetrical initial positions, symmetrical destination positions, and symmetrical situations. These symmetric features are further conducive to quantitative analysis of the pedestrian behaviors.

\subsection{Experiment setup}

Our circle antipode experiments are organized on a square of Beijing Jiaotong University in December 2017, and in all 64 participants took part in the experiments \citep{Xiao2019}. Two types of circles ($r$ = 5 m and $r$ = 10 m) are involved in the experiments, while the corresponding pedestrian starting/destination locations are respectively labeled with conspicuous number marks on the ground.

Here, we carried out 8 types of experiments with different sizes of circles and different numbers of participants, and each type of experiments repeated 4 times in total. For the sake of convenience in description, the circle antipode experiment with 5 m circle and 8 participants is expressed as 5m-8p experiment. The experiments contain four different numbers of participants (8, 16, 32, 64), and these numbers are set to ensure the efficiency of experiments and the symmetric features of pedestrians. Note that the participants are randomly divided into groups A (1-32) and group B (33-64) before the experiments, and each group contains 32 individuals.

In the warm-up period of the experiments, the participants were suggested to reach the destination as fast as possible, and they are also reminded of the potential safety issues. During the formal experiment stage, the participants were not required to stay inside the circle but must move within a square cordon area outside the circle. It's noted that the cordon area is set to limit the movement scope of participants and prevent the influence of passersby. Also, more detailed experiment processes can be found on the website \url{http://pedynamic.com/circle-antipode-experiments}.

\subsection{Data collection and extraction}

\begin{figure}[ht]
\centering{ \includegraphics[width=1\textwidth]{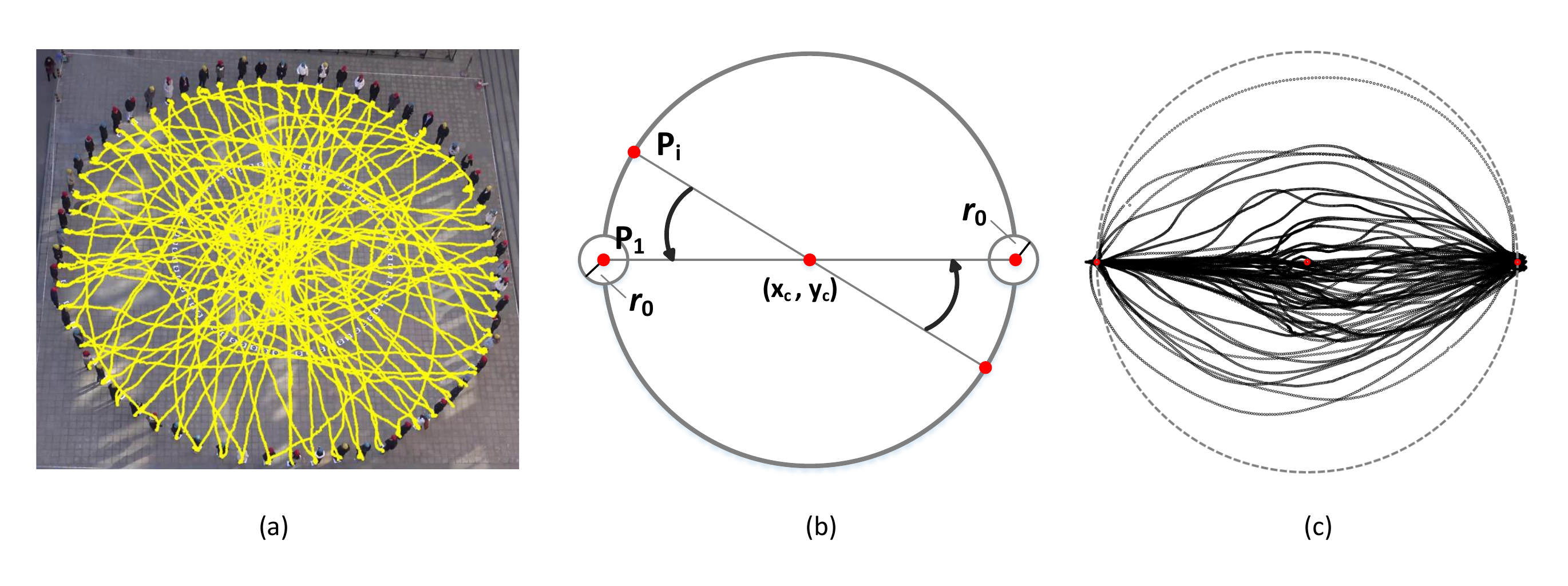}}
\caption{Rotation of the original trajectories in circle antipode experiments.} \label{figRotation}
\end{figure}

% camera 
A high-definition camera was placed on the top of the building beside the ground, and it recorded the entire movements of pedestrians in the whole experiment. Based on the videos, the color mode of PeTrack software \cite{Boltes2013,Boltes2010} was applied to extract pedestrian movement data. It is noted that different colors of caps (i.e., blue, red, and yellow) have been respectively assigned to the pedestrians with different heights (i.e., under 165 cm, 165 cm - 175 cm, and above 175 cm) before the formal experiments. Besides, the pedestrians were requested to avoid wear similar colors to the caps to ensure the effectiveness of recognition.

% 轨迹提取和使用
The pedestrian trajectories which are presented as the highlighted lines in Fig. \ref{figRotation}-a are the core data extracted from these videos, and these data also proved a basis for the following quantitative exploration of the pedestrian experiments.

% Trajectories 
In the experiments, the original trajectory is denoted according to the pedestrian index $i$ and time step $t$, 
\begin{equation}\label{eq trajectories original}
\bm{s}_i(t) =\{(x_i(t), \, y_i(t)) \ | \ 1 \leq i \leq N,\ t_i^{start} \leq t \leq t_i^{dest}, \ i,\, t\, \in \mathbb{Z} \}, 
\end{equation}
where $x_i(t)$ and $y_i(t)$ are the position of trajectories along x-axis and y-axis, respectively. $N$ represents the pedestrian count participating in the corresponding experiment. $t_i^{start}$ refers to the departure time from the cut-off circle around the starting position, while $t_i^{dest}$ stands for the arrival time to the cut-off circle around the destination position.

% Trajectory rotation
To facilitate a further investigation of the pedestrian trajectories, the symmetric characteristics of pedestrians in the circle antipode experiments are applied. The original trajectories $\bm{s}_i(t)$ of pedestrian $P_i$ at time $t$ are rotated according to, 
\begin{equation}	\label{eq rotation}
\left \{
\begin{array}{r}
x_i^R (t) = x^c + (x_i (t)-x^c) \cos ( \frac {2\pi(i-1)} {N}) - (y_i (t) - y^c) \sin( \frac {2\pi(i-1)} {N}) \\
y_i^R (t) = y^c + (x_i (t)-x^c) \sin ( \frac {2\pi(i-1)} {N}) + (y_i (t) - y^c) \cos( \frac {2\pi(i-1)} {N}) \\
\end{array}
\right. .
\end{equation}
where $(x^c, y^c)$ is the position of the circle center. Through the trajectory rotation (Fig. \ref{figRotation}), the starting positions (destination positions) of pedestrians converge on the same point. In this respect, the rotated trajectory of pedestrian $P_i$ at time $t$ are given as,
\begin{equation}\label{eq trajectories rotated}
\bm{s}_i^R(t)=\{(x_i^R(t), \, y_i^R(t)) \ | \ 1 \leq i \leq N,\ t_i^{start} \leq t \leq t_i^{dest}, \ i,\, t\, \in \mathbb{Z} \}, 
\end{equation}

\section{Experiment analysis}\label{section 3}

The extracted pedestrian trajectories lay the foundation of quantitative analysis in the experiments, and the symmetric characteristic eliminates the disadvantages of unequal comparison among individuals. In the section, the side preference behavior in circle antipode experiments, which is also a common behavior in pedestrian crowds, is further explored. Generally speaking, the side preference indicates that an individual is likely to select a preferred side to avoid existing or potential conflicts on the way forward.

\subsection{Side preference definition and results}

\begin{figure}[!ht]
\centering 
%\hspace{-0ex} \vspace{-1ex}
{ \includegraphics[width=1\textwidth]{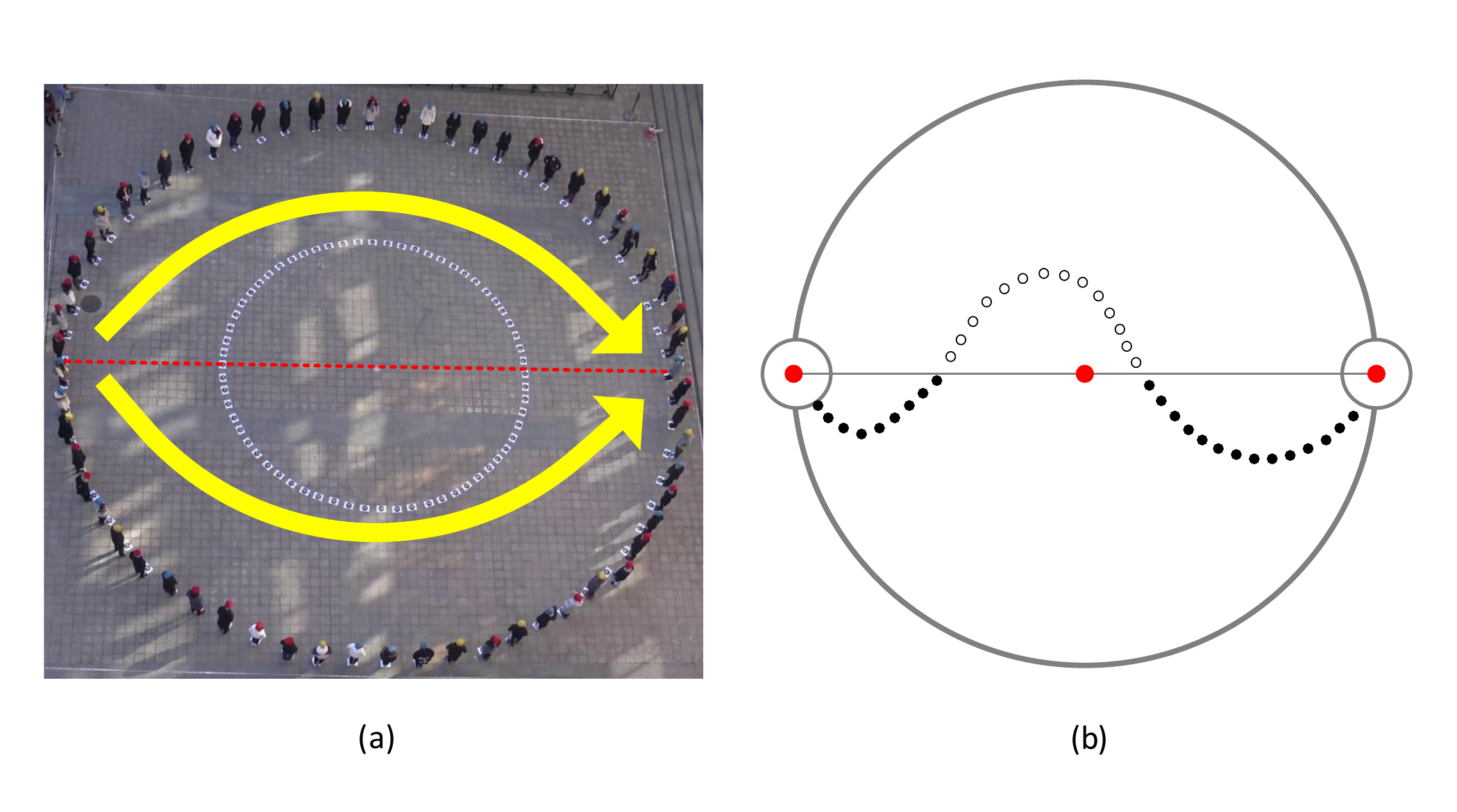}}
\caption{Sketch maps of side preference behavior.} \label{figsidesketch}
\end{figure}

During the circle antipode experiments, a complicated conflicting area is formulated in the center region, and the pedestrians need to choose a preferred side. In the experiments, the side choice is acquired according to the semi-circle choice of a pedestrian as shown in Fig. \ref{figsidesketch}-a. To further quantitatively determine the side choice of a pedestrian, the trajectories in the left and right semi-circles are respectively counted and compared (Fig. \ref{figsidesketch}-b). The side preference of pedestrian $P_i$ in the experiments are determined according to,
\begin{equation}	\label{eq side preferred}
z_i = \left \{ 
\begin{array}{cc}
\rm{Rigth \ Side \ Preferred}, & card(L_i) \leq card(R_i) \\
\rm{Left \ Side \ Preferred}, & card(L_i) > card(R_i) \\
\end{array}
\right. .
\end{equation}
where $card(L_i)$ and $card(R_i)$ respectively indicates the number of the collection of left side trajectories and right side trajectories for pedestrian $P_i$. Suggest that the circle center $(x^c, y^c)$ is the origin of coordinate, the side characteristics of rotated trajectory point $s_i(t)$ is judged by,
\begin{equation}	\label{eq side trajectory}
s_i(t) \in \left \{ 
\begin{array}{cc}
L_i, & y_i(t) \leq 0 \\
R_i, & y_i(t) > 0 \\
\end{array}
\right. .
\end{equation}

Accordingly, the side preferences of 64 pedestrians in the 32 experiments are calculated, and the results are shown in Table. \ref{tab2}. In the table, the row numbers from 1 to 32 represent the experiment indexes, and the column numbers from 1 to 64 represent the pedestrian indexes. Besides, the right side preferred pedestrian is denoted with a solid point $\bullet$, while the left side preferred pedestrian is a hollow point $\circ$. Overall, 960 pedestrian samples are involved in our experiments.

\begin{table}
\renewcommand\arraystretch{0.5}
\centering
\caption{Side preference of individuals in the experiments.} \label{tab2}
\begin{tabular}{lp{0.1cm}<{\centering}p{0.1cm}<{\centering}p{0.1cm}<{\centering}p{0.1cm}<{\centering}p{0.1cm}<{\centering}p{0.1cm}<{\centering}p{0.1cm}<{\centering}p{0.1cm}<{\centering}p{0.1cm}<{\centering}p{0.1cm}<{\centering}p{0.1cm}<{\centering}p{0.1cm}<{\centering}p{0.1cm}<{\centering}p{0.1cm}<{\centering}p{0.1cm}<{\centering}p{0.1cm}<{\centering}p{0.1cm}<{\centering}p{0.1cm}<{\centering}p{0.1cm}<{\centering}p{0.1cm}<{\centering}p{0.1cm}<{\centering}p{0.1cm}<{\centering}p{0.1cm}<{\centering}p{0.1cm}<{\centering}p{0.1cm}<{\centering}p{0.1cm}<{\centering}p{0.1cm}<{\centering}p{0.1cm}<{\centering}p{0.1cm}<{\centering}p{0.1cm}<{\centering}p{0.1cm}<{\centering}p{0.1cm}<{\centering}}
\toprule
	&1&2&3&4&5&6&7&8&9&10&11&12&13&14&15&16&17&18&19&20&21&22&23&24&25&26&27&28&29&30&31&32\\ 
\midrule
1&$\bullet$&$\bullet$&$\bullet$&$\bullet$&$\bullet$&$\bullet$&$\bullet$&$\bullet$& & & & & & &$\bullet$&$\bullet$&$\bullet$&$\bullet$&$\bullet$&$\circ$& & & & & & & & & & &$\bullet$&$\bullet$\\
2&$\bullet$&$\bullet$& & & & & & &$\bullet$&$\bullet$&$\bullet$&$\circ$&$\bullet$&$\bullet$&$\bullet$&$\bullet$&$\bullet$&$\bullet$& & & & & & & & & & &$\circ$&$\circ$&$\bullet$&$\bullet$\\
3&$\bullet$&$\circ$&$\bullet$&$\circ$& & & & & & & & & & &$\bullet$&$\circ$&$\circ$&$\circ$&$\bullet$&$\circ$&$\circ$&$\bullet$&$\circ$&$\bullet$& & & & & & &$\bullet$&$\bullet$\\
4&$\bullet$&$\bullet$& & & & & & & & & & &$\bullet$&$\circ$&$\bullet$&$\bullet$&$\bullet$&$\bullet$& & & & & & &$\circ$&$\circ$&$\circ$&$\bullet$&$\bullet$&$\bullet$&$\bullet$&$\bullet$\\
5&$\bullet$&$\bullet$&$\circ$&$\bullet$&$\bullet$&$\circ$& & & & & & & & &$\bullet$&$\circ$&$\bullet$&$\circ$&$\bullet$&$\bullet$& & & & & & & & & & &$\circ$&$\circ$\\
6&$\bullet$&$\circ$& & & & & & & & &$\circ$&$\bullet$&$\bullet$&$\circ$&$\bullet$&$\bullet$&$\bullet$&$\circ$& & & & & & & & & & &$\circ$&$\bullet$&$\bullet$&$\circ$\\
7&$\bullet$&$\circ$&$\bullet$&$\bullet$& & & & & & & & & & &$\bullet$&$\bullet$&$\circ$&$\bullet$&$\bullet$&$\bullet$&$\circ$&$\bullet$& & & & & & & & &$\bullet$&$\bullet$\\
8&$\bullet$&$\circ$& & & & & & & & & & &$\bullet$&$\circ$&$\bullet$&$\circ$&$\bullet$&$\circ$& & & & & & & & &$\bullet$&$\bullet$&$\circ$&$\bullet$&$\bullet$&$\bullet$\\
9&$\bullet$&$\circ$&$\circ$&$\bullet$&$\circ$&$\circ$&$\circ$&$\bullet$& & & & & & &$\bullet$&$\circ$&$\circ$&$\circ$&$\bullet$&$\circ$& & & & & & & & & & &$\bullet$&$\circ$\\
10&$\circ$&$\circ$& &$\bullet$& & & & &$\circ$&$\circ$&$\circ$&$\circ$&$\circ$&$\circ$&$\circ$&$\circ$&$\circ$&$\circ$& & & & & & & & & & &$\circ$&$\circ$&$\circ$&$\bullet$\\
11&$\bullet$&$\circ$&$\bullet$& & & & & & & & & & & &$\bullet$&$\bullet$&$\bullet$&$\bullet$&$\bullet$&$\bullet$&$\bullet$&$\bullet$&$\bullet$&$\bullet$& & & & & & &$\bullet$&$\bullet$\\
12&$\circ$&$\bullet$& & & &$\bullet$& & & & & & &$\bullet$&$\bullet$&$\circ$&$\circ$&$\circ$&$\circ$& & & & & & &$\bullet$&$\bullet$&$\bullet$&$\bullet$&$\bullet$&$\bullet$&$\circ$&$\circ$\\
13&$\circ$&$\circ$&$\circ$&$\bullet$&$\circ$& & & & & & & & & &$\circ$&$\bullet$&$\circ$&$\bullet$&$\circ$&$\bullet$& & & & & & & & & & &$\circ$&$\bullet$\\
14&$\bullet$&$\circ$& & & & & & & & &$\circ$&$\bullet$&$\bullet$&$\bullet$&$\circ$&$\bullet$&$\circ$&$\bullet$& & & & & & & & & & &$\circ$&$\circ$&$\circ$&$\bullet$\\
15&$\circ$&$\circ$&$\bullet$&$\circ$& & & & & & & & & & &$\bullet$&$\circ$&$\bullet$&$\bullet$&$\circ$&$\bullet$&$\bullet$&$\circ$& & & & & & & & &$\circ$&$\circ$\\
16&$\circ$&$\circ$& & & & & & & & & & &$\bullet$&$\bullet$&$\bullet$&$\circ$&$\bullet$&$\circ$& & & & & & & & &$\bullet$&$\bullet$&$\bullet$&$\bullet$&$\bullet$&$\bullet$\\
17&$\circ$&$\bullet$&$\circ$&$\bullet$&$\bullet$&$\bullet$&$\circ$&$\circ$& & & & & & &$\circ$&$\bullet$&$\circ$&$\bullet$&$\circ$&$\bullet$& & & & & & & & & & &$\circ$&$\bullet$\\
18&$\bullet$&$\circ$& & & & & & &$\bullet$&$\bullet$&$\bullet$&$\bullet$&$\bullet$&$\circ$&$\bullet$&$\circ$&$\bullet$&$\circ$& & & & & & & & & & &$\bullet$&$\bullet$&$\bullet$&$\bullet$\\
19&$\circ$&$\bullet$&$\bullet$&$\bullet$& & & & & & & & & & &$\bullet$&$\bullet$&$\bullet$&$\bullet$&$\circ$&$\bullet$&$\bullet$&$\bullet$&$\bullet$&$\bullet$& & & & & & &$\circ$&$\bullet$\\
20&$\bullet$&$\bullet$& & & & & & & & & & &$\bullet$&$\bullet$&$\bullet$&$\bullet$&$\bullet$&$\bullet$& & & & & & &$\bullet$&$\bullet$&$\bullet$&$\bullet$&$\bullet$&$\bullet$&$\circ$&$\bullet$\\
21&$\bullet$&$\bullet$&$\bullet$&$\circ$&$\bullet$&$\circ$& & & & & & & & &$\bullet$&$\bullet$&$\bullet$&$\bullet$&$\bullet$&$\bullet$& & & & & & & & & & &$\bullet$&$\bullet$\\
22&$\bullet$&$\bullet$& & & & & & & & &$\bullet$&$\bullet$&$\circ$&$\bullet$&$\bullet$&$\bullet$&$\circ$&$\bullet$& & & & & & & & & & &$\circ$&$\bullet$&$\bullet$&$\circ$\\
23&$\bullet$&$\bullet$&$\bullet$&$\bullet$& & & & & & & & & & &$\bullet$&$\bullet$&$\bullet$&$\bullet$&$\bullet$&$\bullet$&$\bullet$&$\bullet$& & & & & & & & &$\bullet$&$\bullet$\\
24&$\bullet$&$\bullet$& & & & & & & & & & &$\bullet$&$\bullet$&$\bullet$&$\circ$&$\bullet$&$\bullet$& & & & & & & & &$\bullet$&$\circ$&$\bullet$&$\bullet$&$\bullet$&$\bullet$\\
25&$\circ$&$\bullet$&$\bullet$&$\bullet$&$\circ$&$\bullet$&$\bullet$&$\bullet$& & & & & & &$\circ$&$\bullet$&$\bullet$&$\bullet$&$\bullet$&$\bullet$& & & & & & & & & & &$\bullet$&$\bullet$\\
26&$\circ$&$\bullet$& & & & & & &$\bullet$&$\circ$&$\bullet$&$\bullet$&$\bullet$&$\bullet$&$\circ$&$\bullet$&$\bullet$&$\circ$& & & & & & & & & & &$\bullet$&$\bullet$&$\circ$&$\bullet$\\
27&$\circ$&$\bullet$&$\circ$&$\bullet$& & & & & & & & & & &$\circ$&$\bullet$&$\bullet$&$\circ$&$\bullet$&$\bullet$&$\circ$&$\bullet$&$\bullet$&$\circ$& & & & & & &$\bullet$&$\bullet$\\
28&$\circ$&$\circ$& & & & & & & & & & &$\bullet$&$\bullet$&$\bullet$&$\bullet$&$\bullet$&$\bullet$& & & & & & &$\bullet$&$\circ$&$\circ$&$\bullet$&$\bullet$&$\circ$&$\bullet$&$\circ$\\
29&$\bullet$&$\bullet$&$\bullet$&$\circ$&$\bullet$&$\bullet$& & & & & & & & &$\bullet$&$\circ$&$\bullet$&$\bullet$&$\bullet$&$\bullet$& & & & & & & & & & &$\bullet$&$\circ$\\
30&$\bullet$&$\bullet$& & & & & & & & &$\bullet$&$\bullet$&$\bullet$&$\bullet$&$\bullet$&$\bullet$&$\bullet$&$\bullet$& & & & & & & & & & &$\bullet$&$\bullet$&$\bullet$&$\bullet$\\
31&$\bullet$&$\circ$&$\bullet$&$\circ$& & & & & & & & & & &$\bullet$&$\circ$&$\bullet$&$\circ$&$\bullet$&$\circ$&$\circ$&$\bullet$& & & & & & & & &$\bullet$&$\circ$\\
32&$\bullet$&$\bullet$& & & & & & & & & & &$\bullet$&$\bullet$&$\bullet$&$\bullet$&$\bullet$&$\bullet$& & & & & & & & &$\circ$&$\circ$&$\bullet$&$\circ$&$\bullet$&$\bullet$\\
33&$\bullet$&$\bullet$&$\bullet$&$\circ$&$\bullet$&$\bullet$&$\bullet$&$\bullet$& & & & & & &$\bullet$&$\bullet$&$\bullet$&$\circ$&$\bullet$&$\circ$& & & & & & & & & & &$\bullet$&$\bullet$\\
34&$\circ$&$\circ$& & & & & & &$\bullet$&$\bullet$&$\bullet$&$\bullet$&$\bullet$&$\circ$&$\bullet$&$\circ$&$\bullet$&$\bullet$& & & & & & & & & & &$\bullet$&$\bullet$&$\bullet$&$\bullet$\\
35&$\circ$&$\bullet$&$\bullet$&$\bullet$& & & & & & & & & & &$\bullet$&$\bullet$&$\circ$&$\bullet$&$\circ$&$\circ$&$\bullet$&$\bullet$&$\circ$&$\circ$& & & & & & &$\bullet$&$\bullet$\\
36&$\circ$&$\bullet$& & & & & & & & & & &$\bullet$&$\circ$&$\bullet$&$\bullet$&$\bullet$&$\bullet$& & & & & & &$\bullet$&$\bullet$&$\bullet$&$\bullet$&$\circ$&$\bullet$&$\bullet$&$\bullet$\\
37&$\circ$&$\circ$&$\bullet$&$\bullet$&$\circ$&$\bullet$& & & & & & & & &$\circ$&$\bullet$&$\bullet$&$\bullet$&$\bullet$&$\circ$& & & & & & & & & & &$\bullet$&$\bullet$\\
38&$\bullet$&$\circ$& & & & & & & & &$\bullet$&$\bullet$&$\bullet$&$\circ$&$\bullet$&$\bullet$&$\bullet$&$\circ$& & & & & & & & & & &$\bullet$&$\bullet$&$\bullet$&$\bullet$\\
39&$\bullet$&$\circ$&$\bullet$&$\bullet$& & & & & & & & & & &$\bullet$&$\bullet$&$\bullet$&$\bullet$&$\bullet$&$\bullet$&$\circ$&$\bullet$& & & & & & & & &$\bullet$&$\bullet$\\
40&$\circ$&$\bullet$& & & & & & & & & & &$\circ$&$\circ$&$\circ$&$\circ$&$\bullet$&$\bullet$& & & & & & & & &$\bullet$&$\bullet$&$\bullet$&$\bullet$&$\bullet$&$\bullet$\\
41&$\bullet$&$\bullet$&$\bullet$&$\bullet$&$\bullet$&$\bullet$&$\bullet$&$\bullet$& & & & & & &$\bullet$&$\bullet$&$\bullet$&$\bullet$&$\bullet$&$\bullet$& & & & & & & & & & &$\bullet$&$\bullet$\\
42&$\bullet$&$\bullet$& & & & & & &$\bullet$&$\bullet$&$\bullet$&$\bullet$&$\bullet$&$\bullet$&$\bullet$&$\bullet$&$\bullet$&$\bullet$& & & & & & & & & & &$\bullet$&$\bullet$&$\bullet$&$\bullet$\\
43&$\circ$&$\circ$&$\bullet$&$\bullet$& & & & & & & & & & &$\bullet$&$\bullet$&$\bullet$&$\bullet$&$\bullet$&$\bullet$&$\bullet$&$\bullet$&$\bullet$&$\bullet$& & & & & & &$\bullet$&$\bullet$\\
44&$\circ$&$\circ$& & & & & & & & & & &$\circ$&$\circ$&$\circ$&$\circ$&$\circ$&$\circ$& & & & & & &$\bullet$&$\bullet$&$\circ$&$\bullet$&$\circ$&$\circ$&$\circ$&$\circ$\\
45&$\circ$&$\circ$&$\circ$&$\circ$&$\circ$&$\circ$& & & & & & & & &$\circ$&$\circ$&$\circ$&$\circ$&$\circ$&$\circ$& & & & & & & & & & &$\circ$&$\circ$\\
46&$\circ$&$\circ$& & & & & & & & &$\bullet$&$\bullet$&$\bullet$&$\bullet$&$\bullet$&$\bullet$&$\circ$&$\bullet$& & & & & & & & & & &$\bullet$&$\bullet$&$\circ$&$\bullet$\\
47&$\circ$&$\circ$&$\circ$&$\circ$& & & & & & & & & & &$\circ$&$\circ$&$\circ$&$\bullet$&$\circ$&$\circ$&$\circ$&$\circ$& & & & & & & & &$\circ$&$\circ$\\
48&$\circ$&$\bullet$& & & & & & & & & & &$\circ$&$\circ$&$\circ$&$\circ$&$\circ$&$\circ$& & & & & & & & &$\circ$&$\circ$&$\circ$&$\circ$&$\circ$&$\bullet$\\
49&$\bullet$&$\bullet$&$\bullet$&$\bullet$&$\circ$&$\bullet$&$\bullet$&$\bullet$& & & & & & &$\bullet$&$\bullet$&$\bullet$&$\bullet$&$\circ$&$\bullet$& & & & & & & & & & &$\bullet$&$\bullet$\\
50&$\bullet$&$\bullet$& & & & & & &$\bullet$&$\bullet$&$\bullet$&$\bullet$&$\bullet$&$\bullet$&$\bullet$&$\bullet$&$\bullet$&$\bullet$& & & & & & & & & & &$\bullet$&$\bullet$&$\bullet$&$\bullet$\\
51&$\circ$&$\bullet$&$\circ$&$\bullet$& & & & & & & & & & &$\circ$&$\circ$&$\circ$&$\circ$&$\circ$&$\bullet$&$\bullet$&$\circ$&$\bullet$&$\circ$& & & & & & &$\circ$&$\circ$\\
52&$\bullet$&$\bullet$& & & & & & & & & & &$\bullet$&$\bullet$&$\bullet$&$\bullet$&$\bullet$&$\bullet$& & & & & & &$\circ$&$\bullet$&$\bullet$&$\bullet$&$\bullet$&$\bullet$&$\bullet$&$\bullet$\\
53&$\bullet$&$\bullet$&$\bullet$&$\bullet$&$\bullet$&$\bullet$& & & & & & & & &$\bullet$&$\bullet$&$\bullet$&$\bullet$&$\bullet$&$\bullet$& & & & & & & & & & &$\bullet$&$\circ$\\
54&$\circ$&$\circ$& & & & & & & & &$\bullet$&$\bullet$&$\circ$&$\bullet$&$\circ$&$\circ$&$\circ$&$\circ$& & & & & & & & & & &$\circ$&$\circ$&$\circ$&$\circ$\\
55&$\circ$&$\bullet$&$\bullet$&$\bullet$& & & & & & & & & & &$\bullet$&$\bullet$&$\bullet$&$\bullet$&$\bullet$&$\bullet$&$\bullet$&$\bullet$& & & & & & & & &$\circ$&$\bullet$\\
56&$\bullet$&$\bullet$& & & & & & & & & & &$\bullet$&$\bullet$&$\bullet$&$\bullet$&$\bullet$&$\circ$& & & & & & & & &$\bullet$&$\bullet$&$\bullet$&$\bullet$&$\bullet$&$\bullet$\\
57&$\bullet$&$\circ$&$\circ$&$\circ$&$\circ$&$\circ$&$\bullet$&$\bullet$& & & & & & &$\circ$&$\circ$&$\bullet$&$\bullet$&$\circ$&$\circ$& & & & & & & & & & &$\bullet$&$\bullet$\\
58&$\circ$&$\bullet$& & & & & & &$\bullet$&$\bullet$&$\bullet$&$\bullet$&$\bullet$&$\bullet$&$\bullet$&$\bullet$&$\bullet$&$\bullet$& & & & & & & & & & &$\bullet$&$\bullet$&$\bullet$&$\bullet$\\
59&$\circ$&$\bullet$&$\circ$&$\bullet$& & & & & & & & & & &$\circ$&$\bullet$&$\circ$&$\circ$&$\circ$&$\bullet$&$\bullet$&$\bullet$&$\bullet$&$\bullet$& & & & & & &$\bullet$&$\bullet$\\
60&$\bullet$&$\circ$& & & & & & & & & & &$\circ$&$\bullet$&$\bullet$&$\bullet$&$\bullet$&$\circ$& & & & & & &$\bullet$&$\bullet$&$\bullet$&$\bullet$&$\circ$&$\bullet$&$\bullet$&$\bullet$\\
61&$\bullet$&$\circ$&$\bullet$&$\circ$&$\bullet$&$\circ$& & & & & & & & &$\bullet$&$\circ$&$\bullet$&$\circ$&$\bullet$&$\bullet$& & & & & & & & & & &$\bullet$&$\bullet$\\
62&$\bullet$&$\bullet$& & & & & & & & &$\bullet$&$\bullet$&$\bullet$&$\bullet$&$\bullet$&$\bullet$&$\bullet$&$\bullet$& & & & & & & & & & &$\circ$&$\bullet$&$\bullet$&$\bullet$\\
63&$\bullet$&$\bullet$&$\bullet$&$\bullet$& & & & & & & & & & &$\bullet$&$\circ$&$\bullet$&$\bullet$&$\bullet$&$\bullet$&$\circ$&$\bullet$& & & & & & & & &$\bullet$&$\bullet$\\
64&$\bullet$&$\bullet$& & & & & & & & & & &$\bullet$&$\bullet$&$\bullet$&$\circ$&$\bullet$&$\bullet$& & & & & & & & &$\bullet$&$\bullet$&$\bullet$&$\bullet$&$\bullet$&$\bullet$\\
\bottomrule
\end{tabular}
\end{table}

\begin{figure}[!ht]
\centering{ \includegraphics[width=0.8\textwidth]{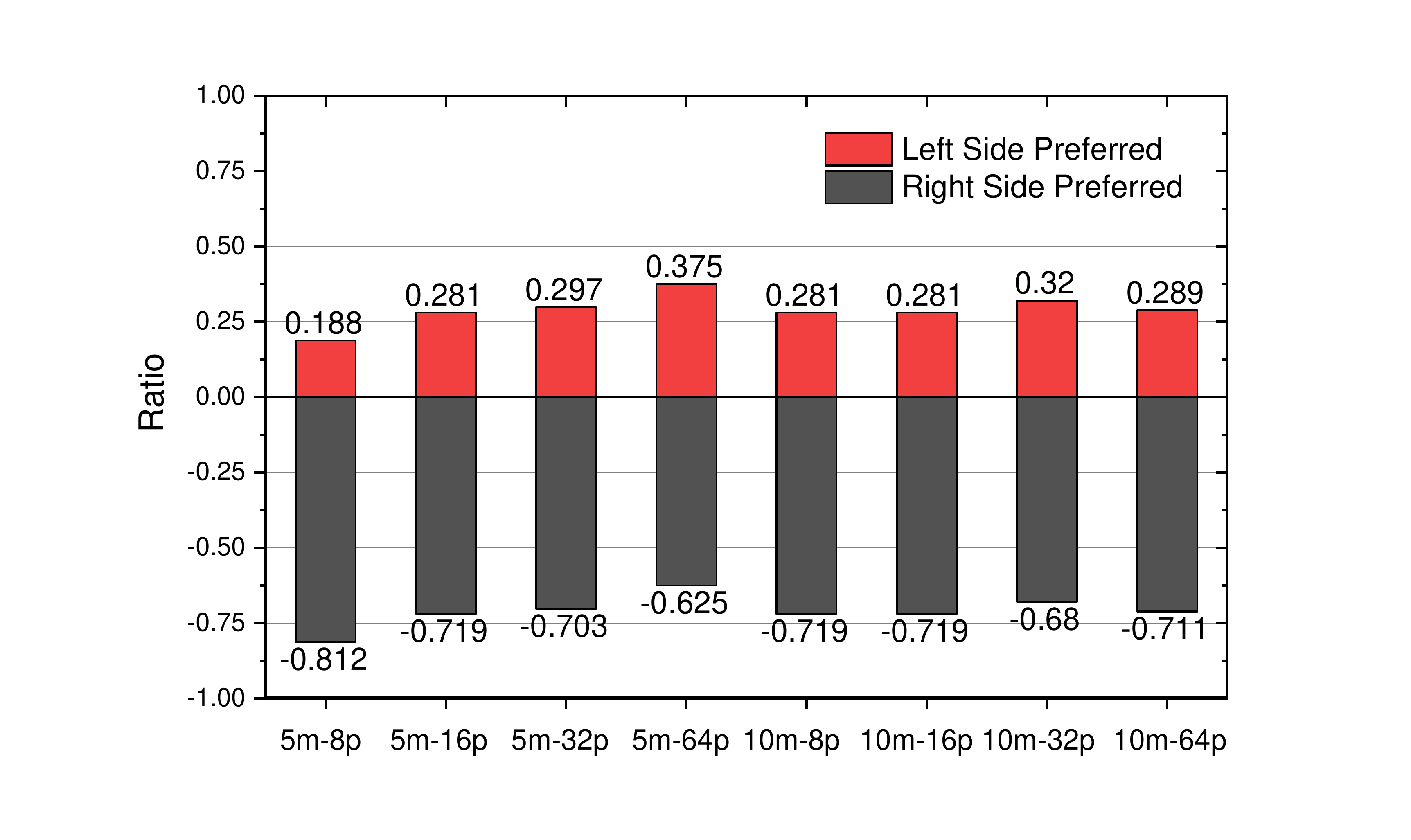}}
\caption{Side preference of experiments.} \label{figSideExperiment}
\end{figure}

Table \ref{tab2} summarizes the pedestrian side preference ratios in different experiments. The results show that 68.75\% of pedestrian samples in all the experiments choose to detour from the right side, and the remaining 31.25\% prefer the left side. From the experiment level (in Fig. \ref{figSideExperiment}), the right side is more preferred no matter in low density situations (e.g., 8p experiments) or in crowded situations (e.g., 64p experiments). It is to say, detouring from the right side is likely to be the customary side choice \citep{Moussaid2009} for the participants, which is almost independent with crowdedness levels. However, the crowdedness levels affect the right side ratios to some extent. For example, in the 5m experiments, fewer pedestrians select the right side under more crowded surroundings. It is because that in a higher density situation pedestrians are more likely to get stuck at the preferred side (i.e. right side), so some of them would change the original side choice and select the other side (i.e. left side). In the 10m experiments, since there is a large enough space for pedestrians moving, the right-right side choice ratio keeps almost stable.

% In a general perspective, 660 individuals choose to detour from the right side and 300 individuals prefer the left side. It is to say, detouring from right side is likely to be the customary side choice \citep{Moussaid2009} for the participants in our experiments. A further investigation is performed from the experiment level, and the pedestrian side preference ratios in different experiments are summarized in Fig. \ref{figSideExperiment}. It is found that the right side is preferred no matter in low density situations (8p experiments) or in crowded situations (64p experiments). It indicates that the qualitative side preference choice pattern of this group of participants is almost fixed, and the specific crowd level just had limited impacts. Averagely, around 72\% of the pedestrians prefer the right side while 28\% of the pedestrians choose the left side in different experiments. Besides, in general, more pedestrians prefer the right side in a low density situation, and the left-right side choice ratios are more balanced in a crowded situation. It is because that in a crowded situation the pedestrians are more likely to get stuck at the preferred side, so some of them would change the original side choice and select the other side.

% 本实验一共有64个人参加，其中32个人参加了14组实验，32个人参加了16组实验。这64个人的side选择被统计出来如图所示。可以看到，个人的side选择呈现了较大的差异性。其中大部分行人都会选择偏向于右侧绕行，而一些行人则呈现出明显的左侧绕行的倾向性。e.g.,ped10,ped45和ped47。
% 而从基于个人的统计分析side选择差异性也可以推断出，行人的side选择是有比较强烈的个人倾向性，虽然整体的右侧选择概率是78而左侧选择概率22，但是个人在多次选择中是会出现明显区别的。

\begin{figure}[!ht]
\centering{ \includegraphics[width=1\textwidth]{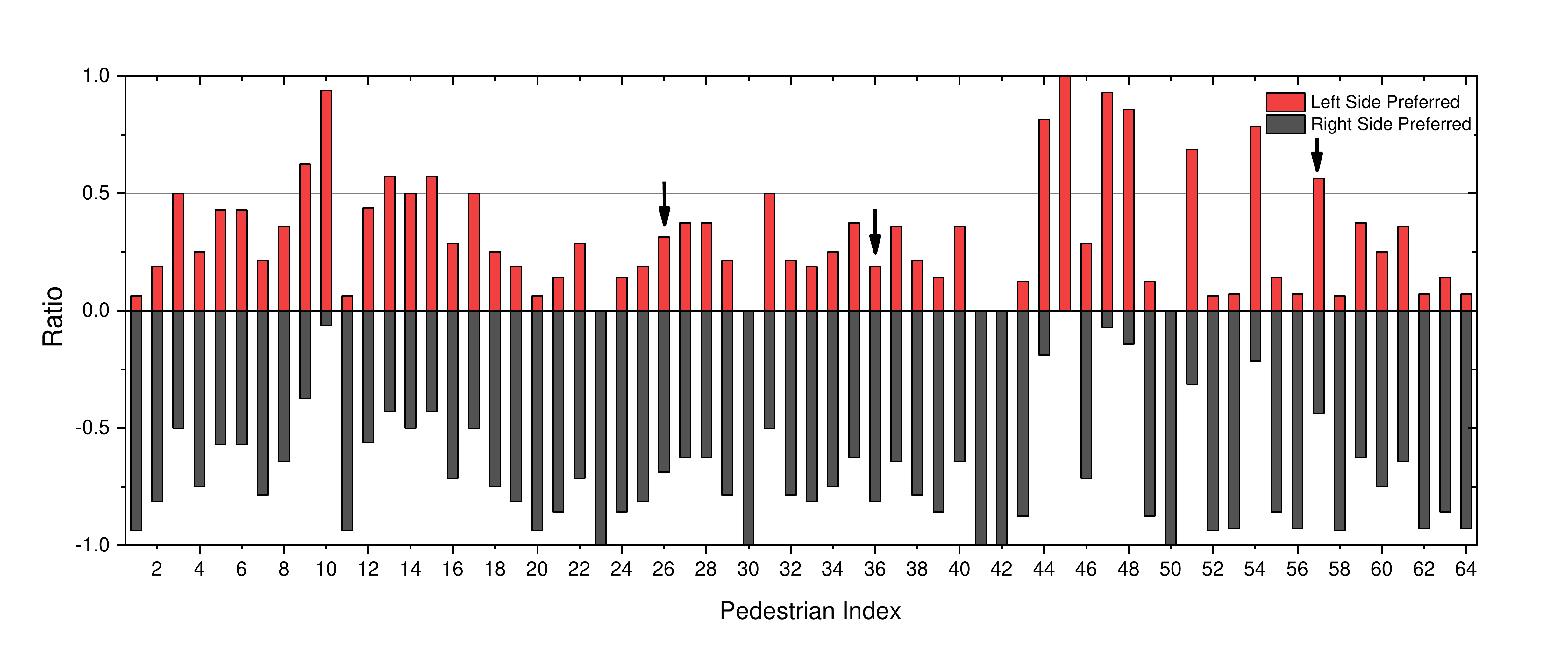}}
\caption{Side preference ratios of individuals.} \label{figsideindividual}
\end{figure}

From an individual perspective, the series of experiments include 64 participants, among whom 32 participants took part in 16 sets of experiments and the other half attended only 14 sets of experiments. Different from the similar side preference ratios among experiments (Fig. \ref{figSideExperiment}), large variances regarding the side preference ratios can be found among the participants, as shown in Fig. \ref{figsideindividual}. Most of the participants prefer to detour from the right side, whereas several ones perform obvious left side preferences, e.g., Pedestrian 45 detour from the left side in all his/her attended experiments.

The repeated experiments demonstrated that the side preferred phenomenon is not caused by the mixture of some pedestrians with definite right side preference and some other with definite left side preference (e.g., 68.75\% of the pedestrians insist on the right side while 31.25\% of pedestrians keep to the left side, so the right side ratio is 68.75\%). Also, the phenomenon is not likely to be a complete stochastic process without pedestrian heterogeneities (e.g., a pedestrian simply preferred to detour from the right side with 68.75\% probability). Since a pedestrian averagely participates in 15 sets of experiments, only 0.36\%(= $0.6875^{15}$) of the pedestrians hang on the right side and only 0.0000027\%(= $0.3125^{15}$) of the pedestrians persist in the left side in all the experiments, theoretically. In our experiments, 5 pedestrians (7.8\%) chose the right side and 1 pedestrian (1.6\%) chose the left side from the beginning to the end, which is obviously different from the theoretical distribution of the stochastic process. Taken as a whole, the side preference of pedestrians should own a more complicated working mechanism and distribution, but not a combination of left side preferred pedestrians and right side preferred pedestrians or a simple stochastic process with heterogeneities.

In summary, each participant owns a particular preference for the side choice, and the aggregation of the various side preferences finally formulates the overall side preference ratio results in our experiments. In the following section, several related factors (i.e., handedness, gender, height) are further included to explore the potential relationship with the participant's side preference.

\subsection{Side preference factor investigation}

\begin{table}[!ht]
\renewcommand\arraystretch{0.7}
\centering
\caption{Participant properties} \label{tab3}
\begin{tabular}{ccccccccp{10cm}} 
\toprule
\multicolumn{4}{c}{Group A} & \multicolumn{4}{c}{Group B}  \\ 
Index & Handedness & Gender & Hat Color & Index & Handedness & Gender & Hat Color\\ 
\midrule
1 & Right & Female & Blue & 33 & Right & Female & Red \\
2 & Right & Female & Blue & 34 & Right & Female & Blue \\
3 & Right & Male & Yellow & 35 & Right & Male & Yellow \\
4 & Right & Female & Red & 36 & Left & Female & Blue \\
5 & Right & Male & Yellow & 37 & Right & Male & Red \\
6 & Right & Female & Blue & 38 & Right & Male & Red \\
7 & Right & Female & Red & 39 & Right & Female & Blue \\
8 & Right & Male & Yellow & 40 & Right & Male & Red \\
9 & Right & Male & Red & 41 & Right & Male & Yellow \\
10 & Right & Male & Red & 42 & Right & Male & Blue \\
11 & Right & Female & Blue & 43 & Right & Female & Red \\
12 & Right & Female & Red & 44 & Right & Female & Yellow \\
13 & Right & Female & Blue & 45 & Right & Male & Red \\
14 & Right & Male & Yellow & 46 & Right & Male & Red \\
15 & Right & Female & Red & 47 & Right & Female & Red \\
16 & Right & Male & Red & 48 & Right & Female & Red \\
17 & Right & Female & Blue & 49 & Right & Male & Red \\
18 & Right & Male & Yellow & 50 & Right & Female & Blue \\
19 & Right & Male & Red & 51 & Right & Female & Blue \\
20 & Right & Male & Red & 52 & Right & Male & Yellow \\
21 & Right & Male & Yellow & 53 & Right & Male & Red \\
22 & Right & Female & Blue & 54 & Right & Male & Red \\
23 & Right & Female & Red & 55 & Right & Female & Blue \\
24 & Right & Male & Yellow & 56 & Right & Male & Yellow \\
25 & Right & Male & Yellow & 57 & Left & Male & Red \\
26 & Left & Female & Blue & 58 & Right & Male & Blue \\
27 & Right & Female & Red & 59 & Right & Male & Red \\
28 & Right & Female & Blue & 60 & Right & Male & Red \\
29 & Right & Male & Blue & 61 & Right & Female & Red \\
30 & Right & Female & Blue & 62 & Right & Male & Red \\
31 & Right & Female & Blue & 63 & Right & Male & Yellow \\
32 & Right & Female & Red & 64 & Right & Female & Blue \\
\bottomrule
\end{tabular}
\end{table}

The handedness information of the 64 participants was collected, and only 3 of them (Pedestrian 26, Pedestrian 36 and Pedestrian 57) are left-handers. Indeed the number of left handedness participants is quite limited to draw a solid statistical conclusion, at least there is no obvious side preference tendentiousness and pattern for the three left-hander participants based on the results in Fig. \ref{figsideindividual}.

% The gender information of the participants was obtained as shown in Tab. \ref{tab3}, and it includes 33 males and 31 females. To quantitatively verify the influence of gender in the experiments, the side preference ratios in Tab. \ref{tab3} and Fig. \ref{figsideindividual} are applied to the hypothesis testing. Since the ratio samples in the male and female groups are not able to pass the normality test (the Shapiro-Wilk test is applied here), a non-parametric test is used to test the gender based groups. The Mann–Whitney U test is used to compare the male group and the female group, and the null hypothesis is that a randomly selected value from one sample will be equally less than or greater than a randomly selected value from a second sample. The p-value of the Mann-Whitney U test is 0.157 ($>0.05$), which means that the null hypothesis is accepted and the gender factor does not own a significant impact on the side preference behaviors. 

The influence of the participant’s gender on the side preference is also explored. The gender information of the participants was obtained as shown in Tab. \ref{tab3}, and it includes 33 males and 31 females. We carry out the hypothesis testing based on the participants’ gender information in Tab. \ref{tab3} and individual’s side preference ratio in Fig. \ref{figsideindividual}. Given that the ratio of the sample in the male and female groups are not able to pass the normality test (the Shapiro-Wilk test is applied here), the Mann–Whitney U test, one of the non-parametric tests, is adopted to compare the male group and the female group. In this case, the null hypothesis is that a randomly selected value from one sample will be equally less than or greater than a randomly selected value from the other sample. The result that the p-value of the Mann-Whitney U test is 0.157 ($> 0.05$) reveals the null hypothesis is accepted, and the gender factor does not pose a significant impact on the side preference behaviors.

In the experiments, the hats with different colors (i.e., yellow, red, blue) had been respectively assigned to the participants with different height range (i.e., $> 175cm, 165 cm - 175 cm, < 165 cm$). The detailed assigned results can be found in Tab. \ref{tab3}, and there are 14 yellow hats, 29 red hats, and 21 blue hats. Combined the data in Tab. \ref{tab3} and Fig. \ref{figsideindividual}, the side preference ratios of pedestrians with different height ranges are prepared for hypothesis testing. As the normality test (the Shaprio-Wilk test) is unable to be passed using the current side preferred ratios in different height groups, the no-parametric Kruskal–Wallis H test is considered. Here, the null hypothesis is given as that the three groups of pedestrian side preference ratios have no differences, and the p-value is 0.496 ($>0.05$). It indicates that the null hypothesis is accepted, and the height factor has no significant impact on the pedestrian side preference behaviors. 

\subsection{Consistency and efficiency}

\begin{figure}[!ht]
\centering 
\subfloat[]{ \includegraphics[width=0.5\textwidth]{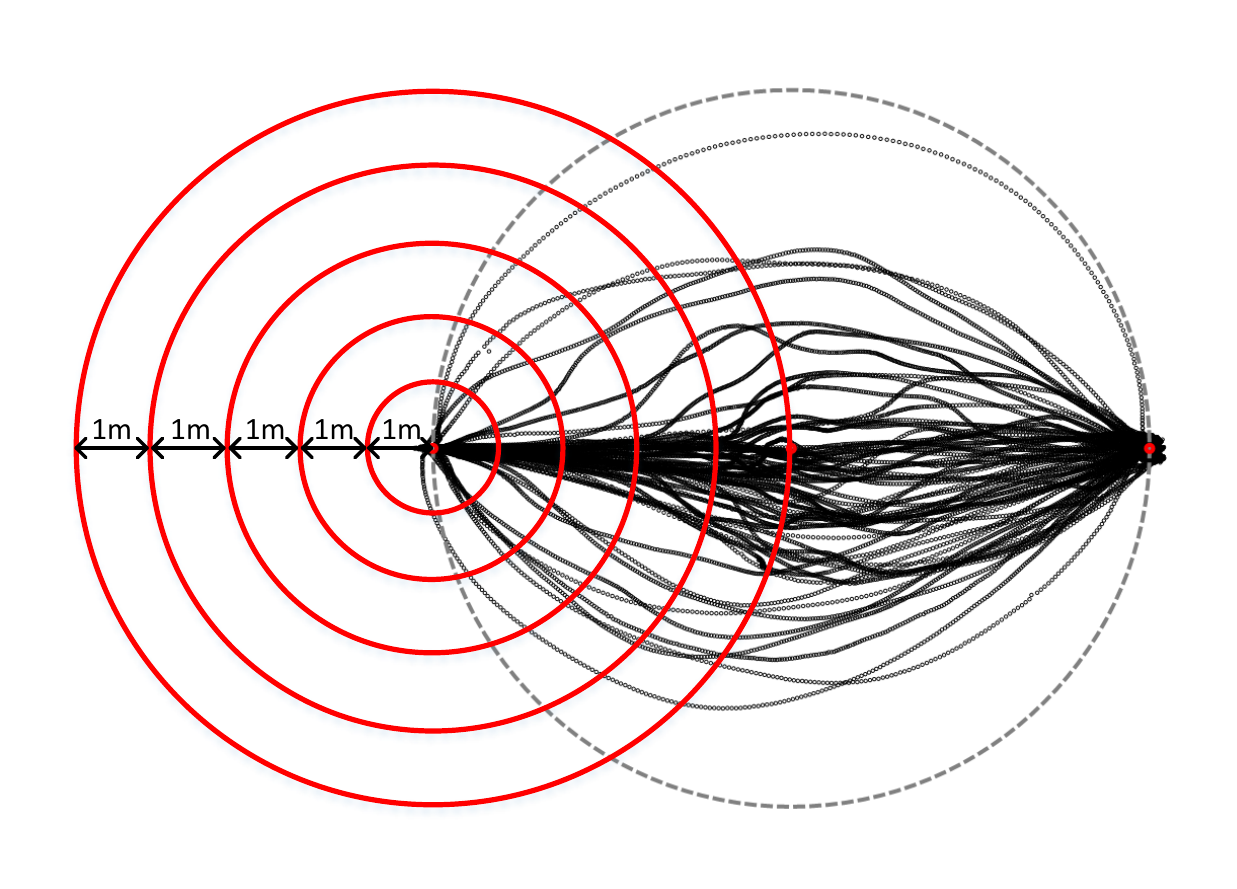}}
\subfloat[]{ \includegraphics[width=0.5\textwidth]{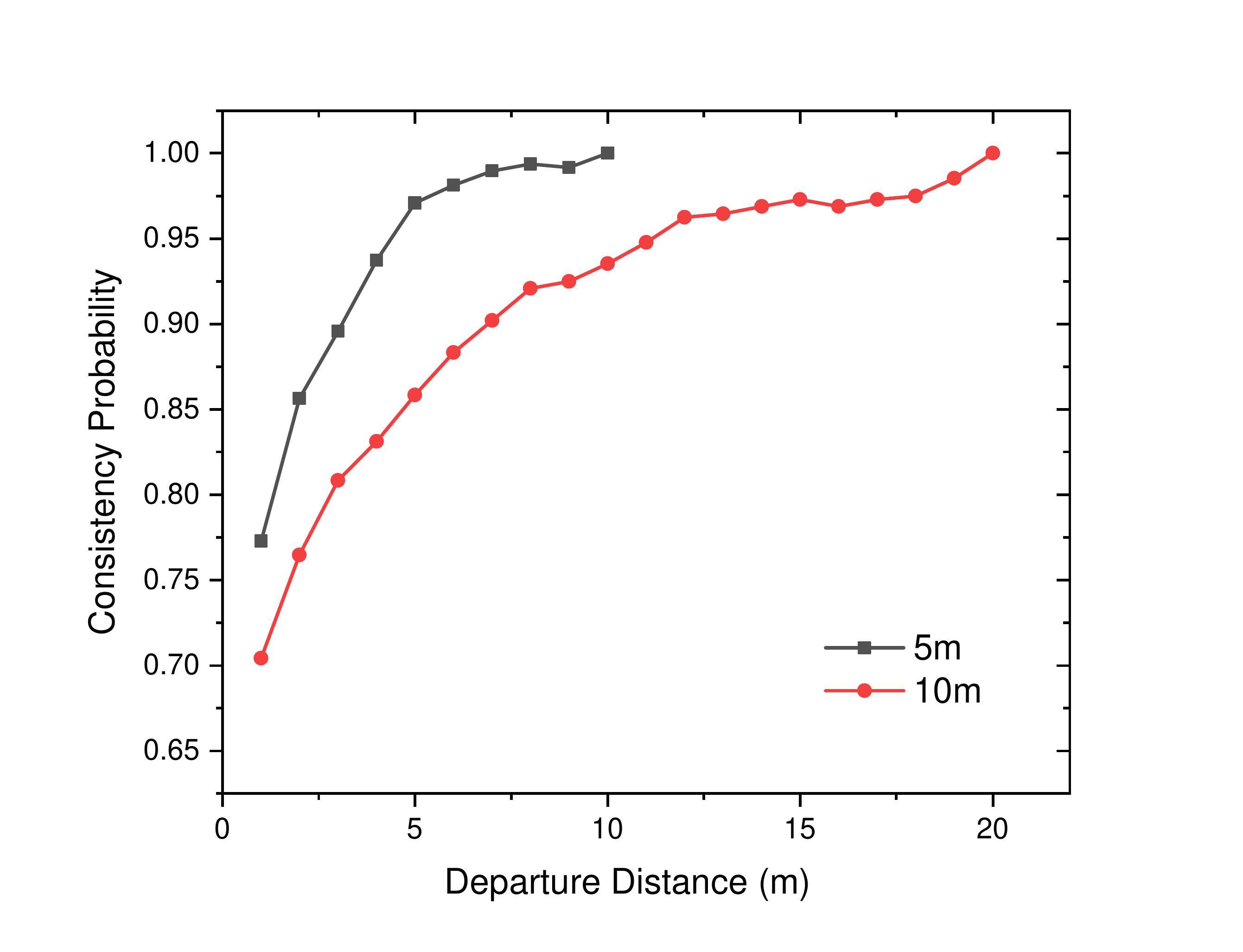}}
\caption{Departure distance consistency of side preference.} \label{figsideconsistency}
\end{figure}

Another study-worthy problem is about the decision moment of side preference behavior in the circle antipode experiments. To address the problem, a general judgment method is proposed based on the rotated trajectories as shown in Fig. \ref{figsideconsistency}-a. The red point in the left represents the starting point, and the red point on the right side represents the destination point. Here, a series of concentric circles are drawn based on the starting point to indicate the pedestrian departure distance from the starting point, and the radii of these circles are respectively 1m, 2m, ... , $n$ m. For the experimental trajectories of each participant, the local side preference inside the different concentric circles are calculated and compared with the global side preference, and the consistency situation of pedestrian $P_i$ is calculated as Eq. \ref{eq consistency}.
\begin{equation}\label{eq consistency}
m_i(d) = \left \{ 
\begin{array}{cc}
1, & {\rm if} \  z_i = z_i^{loc} \\
0, & {\rm otherwise} \\
\end{array}
\right. .
\end{equation}
where $d$ is the radius of the concentric circle and $z_i^{loc}$ is the side preference of pedestrian $P_i$ in the local region. Besides, the overall consistency level can be calculated by Eq.\ref{eq consistency prob}. 
\begin{equation}\label{eq consistency prob}
M(d) = \sum_{i=1}^{n} m_i(d) / n
\end{equation}
Accordingly, the consistency results are shown in Fig. \ref{figsideconsistency}-b. With the growing of departure distance, the level of consistency continuously increases and eventually gets to 1. Besides, the consistency level when the pedestrian just departure from the starting point (e.g., departure distance = 1m) is greater than the theoretic consistency probability which is believed to be about 0.5. In fact, for the 10m experiments, the side preference of over 70\% of pedestrians in the local circle of radius 1m agrees with the global side preference, and the consistency probability is even higher (almost 77.5\%) for the 5m experiments. It indicates that maybe up to 70\% pedestrians in the 10m experiments and about up to 77.5\% pedestrians in the 5m experiments made the side preference choice within the scope of departure distance equaling to 1m. Besides, the consistency probability quickly reaches 90\% at about $1/3$ of the whole distance (3m for the 5m experiments and 7m for the 10m experiments). The results prove that most pedestrians have made their behavioral decisions about which side to walk forward and detour at the very beginning of the motion.

\begin{figure}[!ht]
\centering{ \includegraphics[width=0.8\textwidth]{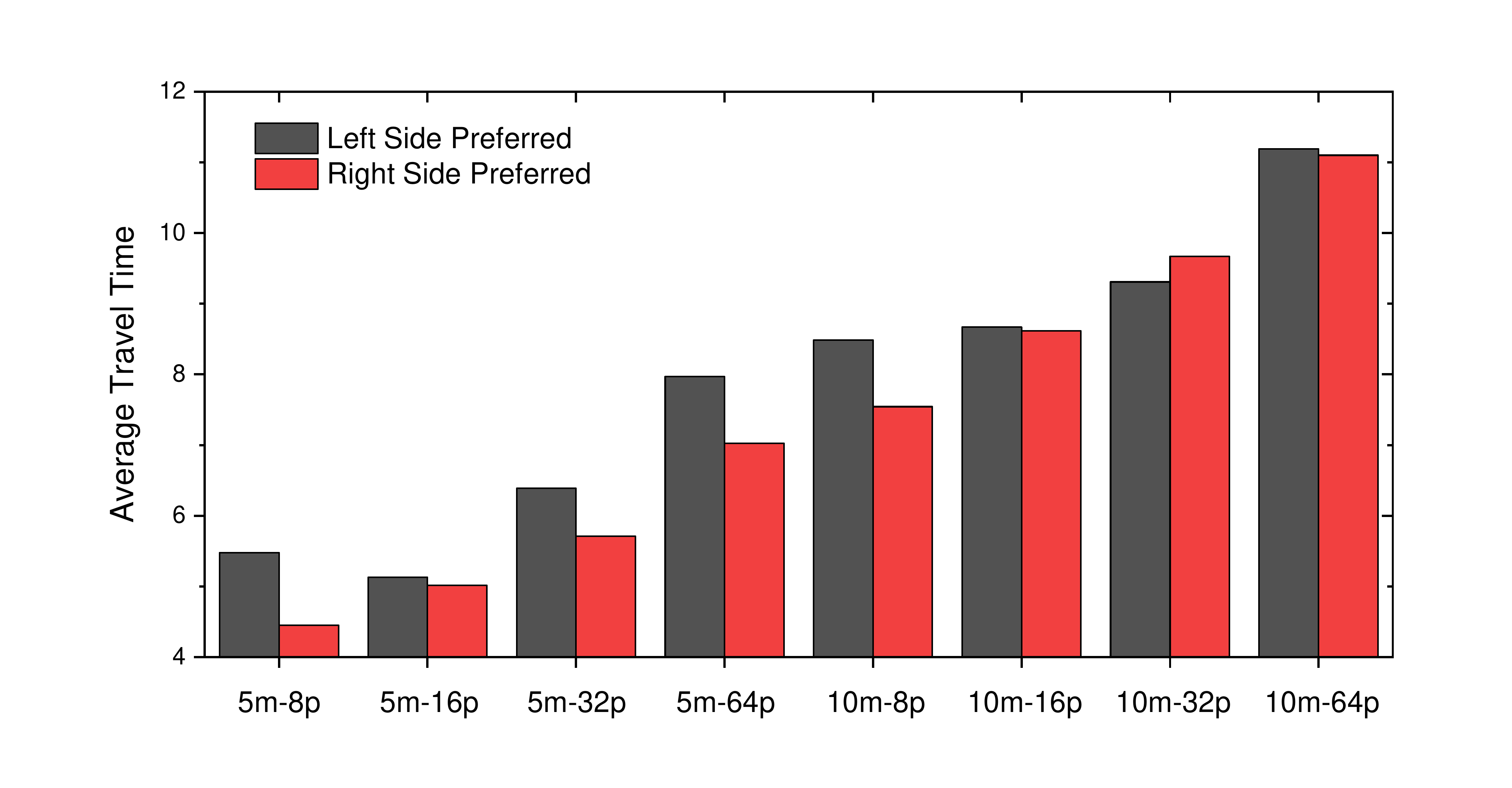}}
\caption{Travel time with different side preference behaviors.} \label{figSideTime}
\end{figure}

% 更进一步的，我们还研究了side选择对运行效率的影响。我们将八组实验中的行人分成了两组即倾向于右侧行走的行人和倾向于左侧行走的行人，并计算了他们的平均旅行时间，如图所示。可以看到，在大部分的实验中（除了10-32p）实验中，倾向于右侧行走的行人都可以更快的到达他们的目的地，并且在5m实验中，这一现象会更为明显。这一现象的产生主要与整个人群的总体运行模式有关，当大部分都朝向右侧绕行运动时，那些朝向左侧的行人一般会遇到更多的来自于其他行人的阻碍和冲突，这些阻碍会影响行人更快的到达目的地。而在5m实验中，由于实验空间更为狭小，行人的避让可能更为困难，所以其他行人的阻碍会对行人运动效率造成更为显著的影响，也因此旅行时间差异会更加明显。

\begin{table}[!ht]
\centering
\caption{Hypothesis test results of travel time for different side preference.} \label{tab4}
\begin{tabular}{ccccccccc}
\toprule
% \multicolumn{5}{c}{5m experiment} \\
 & 5m-8p & 5m-16p & 5m-32p & 5m-64p & 10m-8p & 10m-16p & 10m-32p & 10m-64p \\ 
\midrule
p-value & 0.044 & 0.797 & 0.012 & 0.000 & 0.045 & 0.771 & 0.072 & 0.631 \\
\bottomrule
\end{tabular}
\end{table}

Furthermore, the impact of side preference on the movement efficiency is discussed in the section. The pedestrians in the 8 types of experiments are divided into two groups, i.e., the left side preferred and the right side preferred, and their average travel times are also calculated as shown in Fig. \ref{figSideTime}. To quantitatively test the impact of side preference, the non-parametric Mann-Whitney U test is applied, and the null hypothesis is given as that the side choice has no significant impact on the pedestrian travel time in experiments. The p-value results are found in Tab. \ref{tab4}, and it presents that the side preference shows a significant effect on the travel time in 5m-8p, 5m-32p, 5m-64p, 10m-8p experiments ($p < 0.05$). In these experiments, the statistical results prove that the right side preferred pedestrians can arrive at the destinations in shorter times, and it is believed that a pedestrian choosing the right side would face fewer conflicts in the movement process. It is to say, selecting the dominate side preference (right side in our experiments) is more likely to be a time-saving choice in practical. 

% and the travel time differences are not so obvious according to the test. The reproduction of the differences shall be related to the crowd conflicts. It is known that most pedestrians in the experiments tended to detour from the right side, while those left-side preferred pedestrians are likely to experience more conflicts from others and their travel times to the destination are thus influenced. In the 5m experiments, the limited motion space increases the difficulty of pedestrians avoiding conflicts. Therefore, the impact of other pedestrians might be greater and the difference of travel times are more obvious.  

% It is found that in most of the experiments (except for the 10m-32p experiments), the right side preferred pedestrians would reach their destinations faster, and the phenomena are more obvious in the 5m experiments.

\section{Model simulation}\label{section 4}

\subsection{Voronoi model with side preference behavior} \label{section 4-1}

Famous pedestrian models such as the social force model\citep{Yuan2017} and cellular automata model\citep{Yang2008} have been applied in reproducing the side preference behaviors. In the section, a Voronoi diagram based model\citep{Xiao2016, Qu2018, Xiao2018} is introduced to simulate the pedestrian crowds, especially for the reproduction of the side preference phenomenon. The model mainly applied a geometric concept called Voronoi diagram, which is a kind of basic structure in dividing the space into regions according to a set of points\citep{Voronoi1908, Fortune1987}. Compared with other frequently-used pedestrian models(e.g., Social force model\citep{Helbing1995, Helbing2000}, cellular automaton model\citep{Burstedde2001, Kirchner2002}), the Voronoi diagram based model can well describe the realistic pedestrian behaviors. Actually, the Voronoi diagram owns several special geometric characteristics. First, the Voronoi cell, which contains all the closest region to the related point, is an effective method for defining personal space. Second, the direction to the Voronoi node corresponds to the intermediate space between neighboring pedestrians which is a feasible across or detour behavior, and the direction perpendicular to the Voronoi segment corresponds to the following behavior. 

A core problem of the pedestrian dynamics research is to determine the motion velocity, while the Voronoi model deconstructs the problem into two parts: the direction judgment and the speed calculation \citep{Xiao2016, Qu2018}. In the section, a modified Voronoi model is proposed, and the velocity determination process of pedestrian $P_i$ is reconstructed as shown in Eq. \ref{eq velocity}.
\begin{equation}\label{eq velocity}
\vec{v}_i = \vec{e}_i \cdot v_i
\end{equation}
where $\vec{e}_i$ is a unit vector of velocity of pedestrian $P_i$, and $v_i$ is the size of speed of pedestrian $P_i$. It's noted that a first order velocity formula \citep{Tordeux2014,Tordeux2016} is developed in Eq. \ref{eq velocity}, which means that the acceleration process here would be instantly achieved.

\begin{figure}[!ht]
\centering{\includegraphics[width=0.4\textwidth]{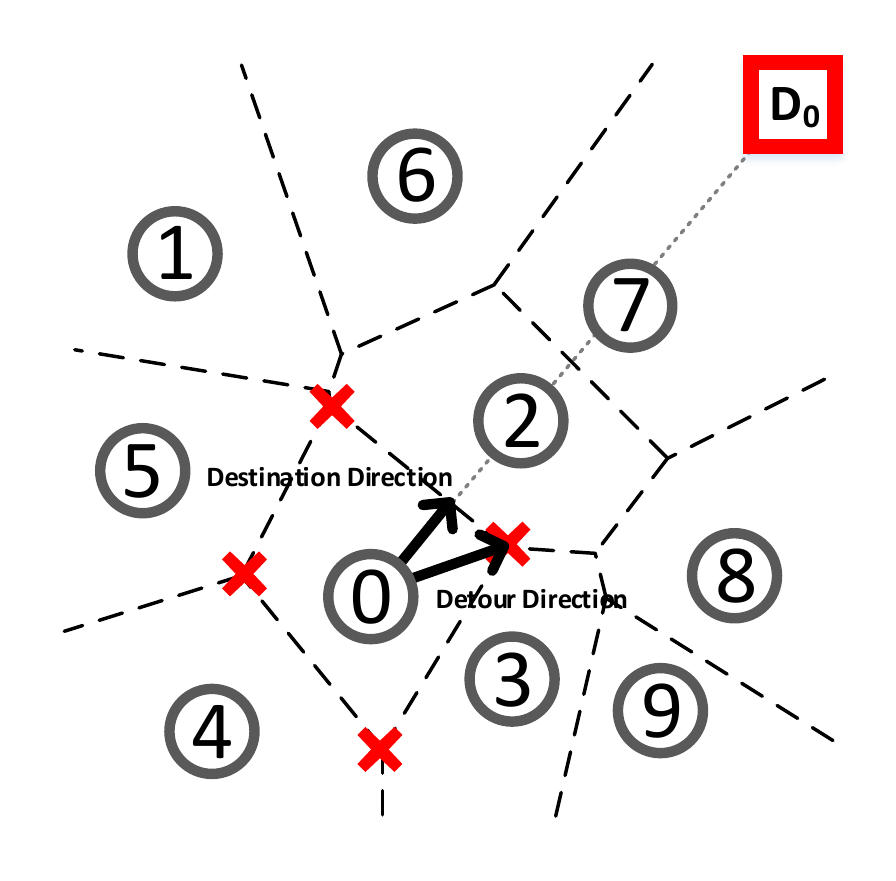}}
\caption{Voronoi diagram based method.} \label{figVSFM}
\end{figure}

The velocity direction in the Voronoi model is related to pedestrian behaviors. Actually, the model in this paper includes two kinds of basic direction choices, i.e., destination direction and detour direction. As seen in Fig.\ref{figVSFM}, the destination direction is the direction to the destination which indicates the forward behaviors, and the detour direction is the direction pointing to the Voronoi node which reveals the detour or across behaviors. At the direction determination stage, the pedestrian makes a choice between destination direction and detour direction according to Eq. \ref{eq velocity direction}. 
\begin{equation}\label{eq velocity direction}
\vec{e}_i=\left\{
\begin{array}{rcl}
\vec{e}^{des}_i, & & {C \geq 0 }\\
\vec{e}^{dtr}_i, & & {\rm{otherwise} }\\
\end{array}.\right.
\end{equation}
Where $\vec{e}^{des}_i$ and $\vec{e}^{dtr}_i$ denote the destination direction and the detour direction, respectively. Note that the destination direction is regarded as a default direction choice for pedestrians, the detour direction will be selected only if the buffer space in the default direction is no longer enough. The buffer space in front is whether enough or not can be found in Eq. \ref{eq direction judgment}
\begin{equation}\label{eq direction judgment}
C = d_{if} - \tau_i (\vec{v}_i - \vec{v}_{if}) \cdot (\vec{l}_{i}-\vec{l}_{if}),
\end{equation}
where $d_{if}$ is the distance between pedestrian $P_i$ and its front pedestrian $P_{if}$. $\tau_i$ is the relaxation time of pedestrian $P_i$. $\vec{v}_i$ and $\vec{v}_{if}$ are respectively the velocity of pedestrian $P_i$ and $P_{if}$. $\vec{l}_i$ and $\vec{l}_{if}$ respectively stands for the location of pedestrian $P_i$ and $P_{if}$. It is noted the front pedestrian $P_{if}$ is defined as the corresponding pedestrian in the neighboring Voronoi cell of the destination direction.

The destination direction $\vec{e}_i^{des}$ is given as Eq. \ref{eq dest direction},
\begin{equation}\label{eq dest direction}
\vec{e}_i^{des} = (\vec{l}_i^{des} - \vec{l}_i)/ \|\vec{l}_i^{des} - \vec{l}_i\|.
\end{equation}
where $l_i^{des}$ refers to the destination of pedestrian $P_i$.

Besides, the detour direction $\vec{e}_i^{dtr}$ is given as Eq.\ref{eq detour direction}.
\begin{equation}\label{eq detour direction}
\vec{e}_i^{dtr} = (\vec{l}_{n^*} - \vec{l}_i)/ \|\vec{l}_{n^*} - \vec{l}_i\|,
\end{equation}
where $\vec{l}_{n^*}$ is the location of the optimal detour choice node among the Voronoi nodes of pedestrian $P_i$. Here, the optimal detour node is determined according to Eq. \ref{eq optimal node}. 
\begin{equation} \label{eq optimal node}
n_i^{*} = \underset{n_{j}\in N_{i}}{\mathrm{\argmax}}(u_1^\alpha \cdot u_2^\beta \cdot u_3^\gamma).
\end{equation}
where $N_i$ denotes the set of Voronoi nodes of pedestrian $P_i$. $\alpha$, $\beta$ and $\gamma$ are three free parameters. In the formula, three factors are taken into consideration for the determination of the optimal detour direction. The first factor is the attraction of destination, that is to say a pedestrian naturally tends to approach the destination rather keep away from it. The tendency of pedestrian is represented as Eq. \ref{eq u1}.
\begin{equation} \label{eq u1}
u_1 = \vec{e}_i^{des}\cdot \vec{e}_{ij}.
\end{equation}

The second factor is the influence of local density. Generally, a pedestrian expects to select those uncrowded routes, and the local density is an excellent index to measure the degree of congestion. In our method\citep{Xiao2018}, the local density of the selected Voronoi node is calculated as the mean value of local densities of its neighboring pedestrians, which is given as Eq. \ref{eq u2}.
\begin{equation} \label{eq u2}
% u_2 = \rho_{n_j} = \sum_{P_k \in T_n_j}^1 \rho_P_k 
u_2 = \rho_{n_j} = \sum_{k=1}^{|T_{n_j}|} \rho_{P_k} 
\end{equation}
where $T_{n_j}$ is the set of the related pedestrians of Voronoi node $n_j$. $\rho_{n_j}$ and $\rho_{P_k}$ are the local density of Voronoi node $n_j$ and pedestrian $P_k$, respectively.

The third factor is regarding the side preference behaviors in pedestrian crowds. Empirical researches \citep{Moussaid2009} shown that the pedestrians usually tend to detour from one side (right/left). The side preference tendency is therefore denoted as Eq. \ref{eq u3}. 
\begin{equation} \label{eq u3}
u_3 = 1+ \vec{e}_i^{des} \times \vec{e}_{ij}
\end{equation}

The speed size of pedestrian $P_i$ is determined according to Eq. \ref{eq velocity size}.
\begin{equation}\label{eq velocity size}
v_i = min(d/\tau_i, v_i^0),
\end{equation}
where $d$ stands for the distance from the pedestrian to the boundary of Voronoi cell in the velocity direction. $v_i^0$ is the desired speed of pedestrian $P_i$.

\subsection{Parameter calibration and simulation results}\label{section 4-2}

In the simulations, it is assumed that the desired speed $v^0 = 1.34 m/s$, the relaxation time $\tau = 0.5 s$ \citep{Helbing1995}. The direction determination parameters $\alpha = 1$, $\beta = -1$. % Besides, regarding the side preference parameter $\gamma$, more details can be found in \ref{section appendix a}.

\begin{figure}[!ht]
\centering{ \includegraphics[width=0.5\textwidth]{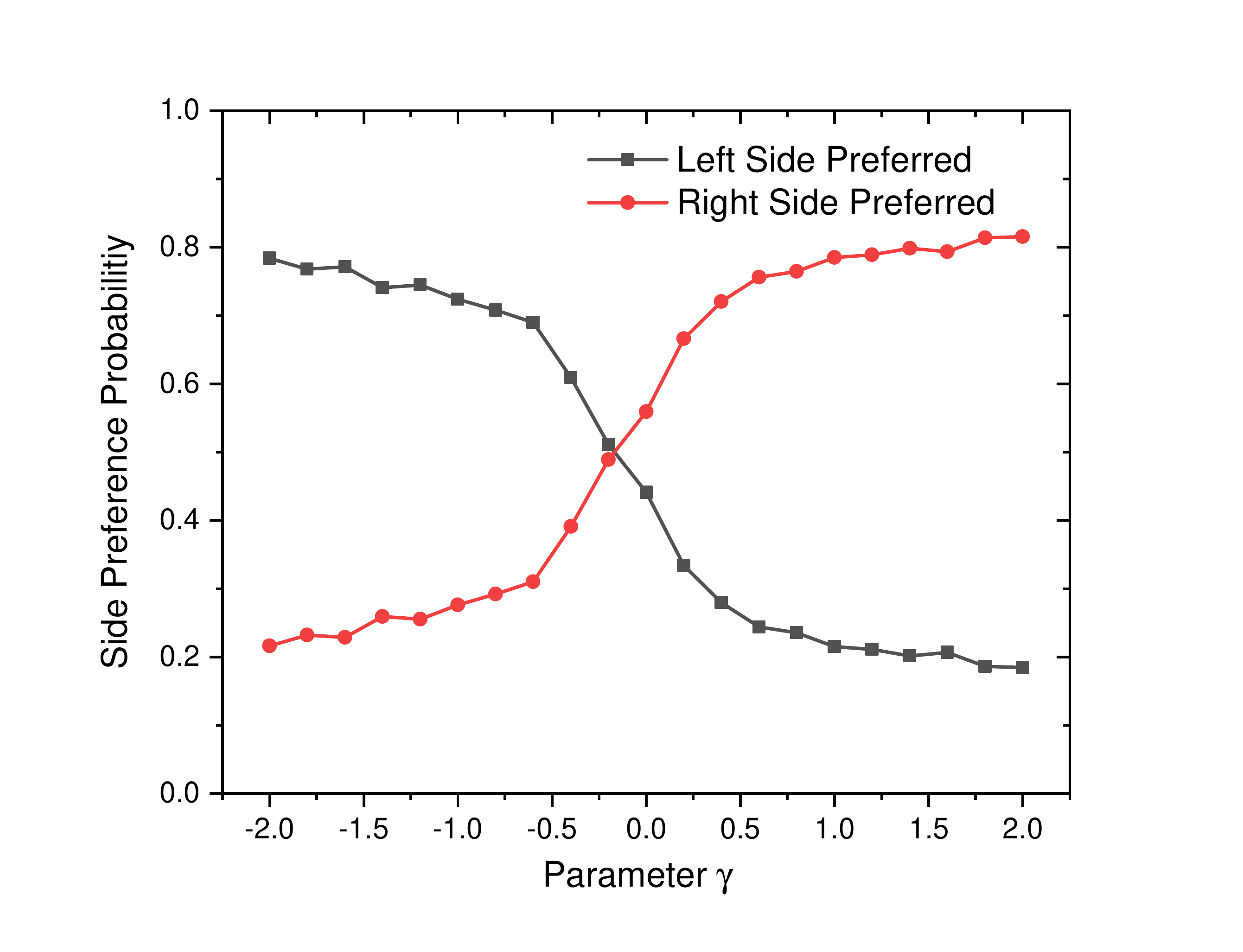}}
\caption{Side preference probability of gamma.} \label{figSidePreferenceProbabilitiy}
\end{figure}

\begin{figure}[!ht]
\subfloat[]{ \includegraphics[width=0.5\textwidth]{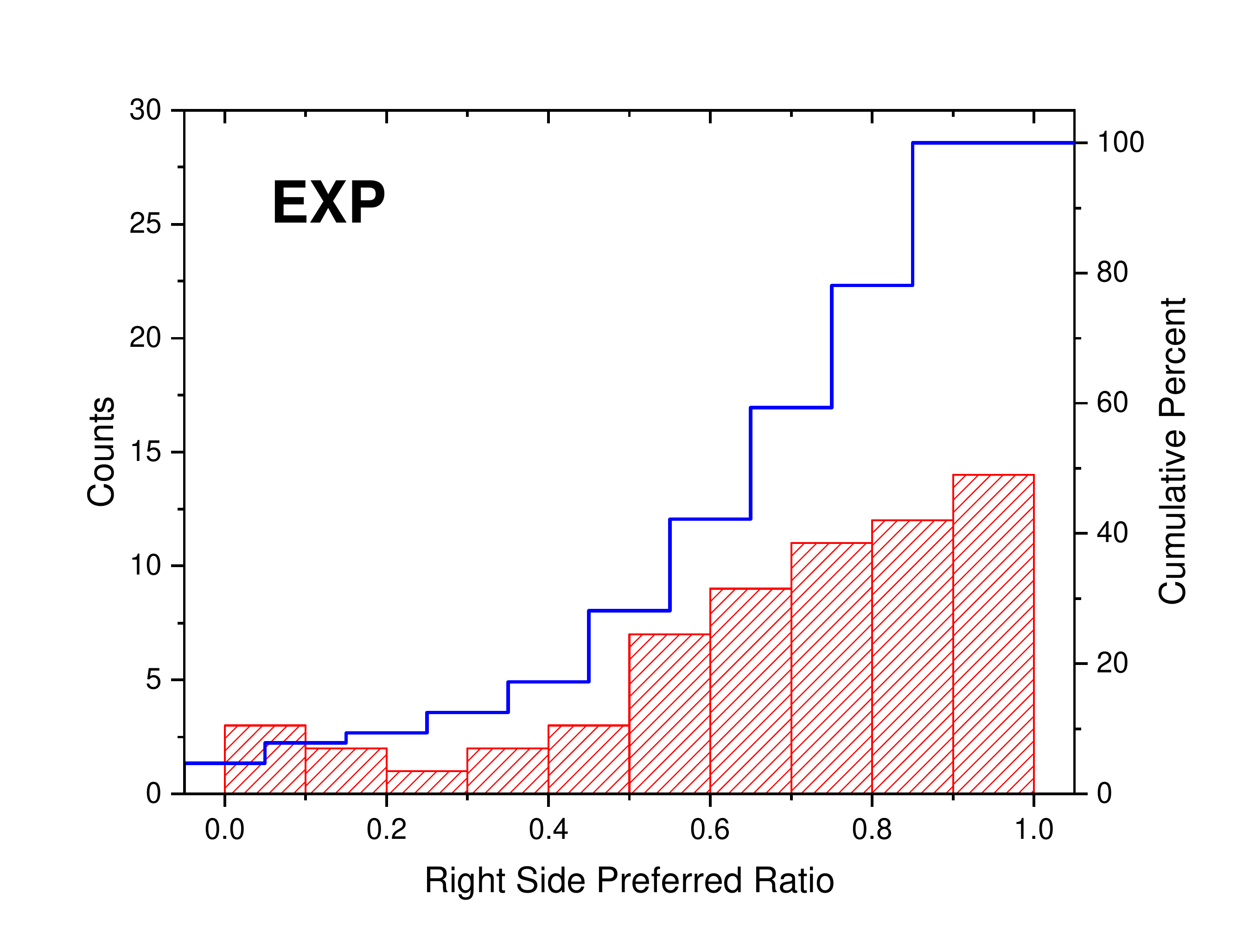}}
\subfloat[]{ \includegraphics[width=0.4\textwidth]{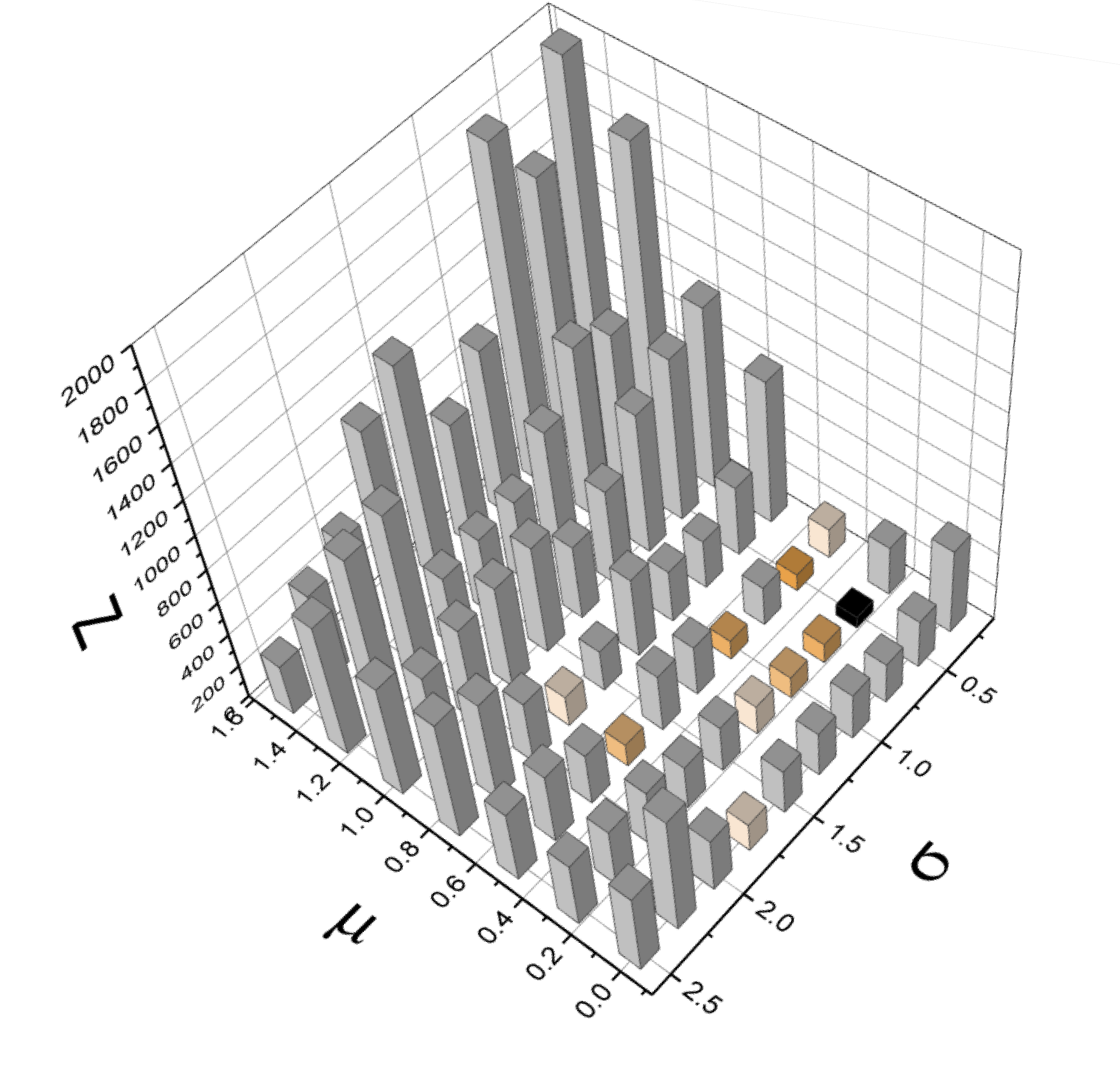}}
\caption{Experiment and calibration results. (a) Right side preferred ratio distribution in experiments. (b) Similarity indexes with different parameters} \label{figsidepercentage}
\end{figure}

In the modified Voronoi model, the parameter $\gamma$ in Eq. \ref{eq velocity direction} is introduced for the reproduction of the practical side preference behaviors. According to the model settings, the side preference behaviors can be transformed with a change of $\gamma$. To reproduce the practical side preference results in experiments, the parameter $\gamma$ is further investigated with different values and distributions.

\begin{figure}[!ht]
\centering
\subfloat[]{ \includegraphics[width=0.4\textwidth]{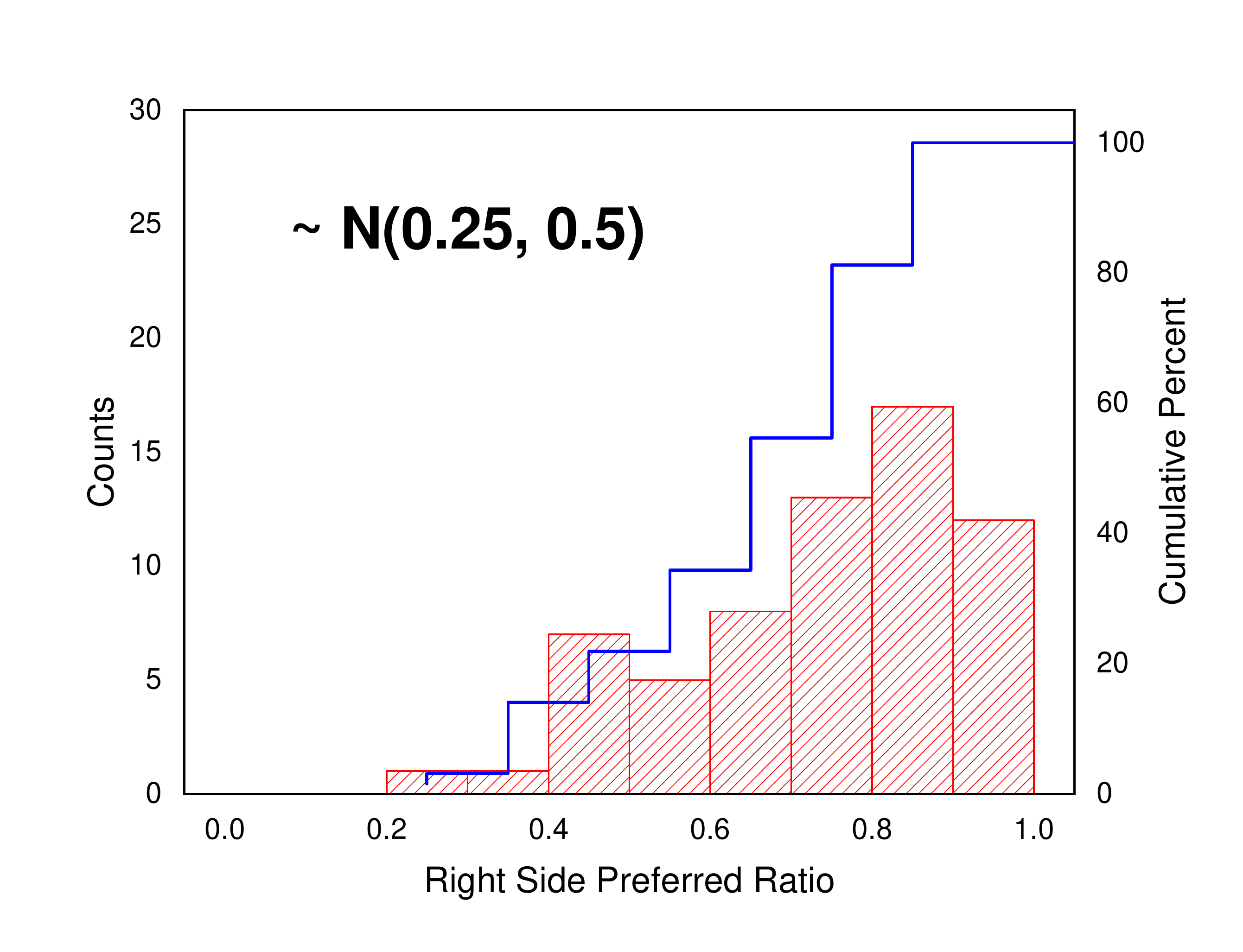}}
\subfloat[] { \includegraphics[width=0.5 \textwidth]{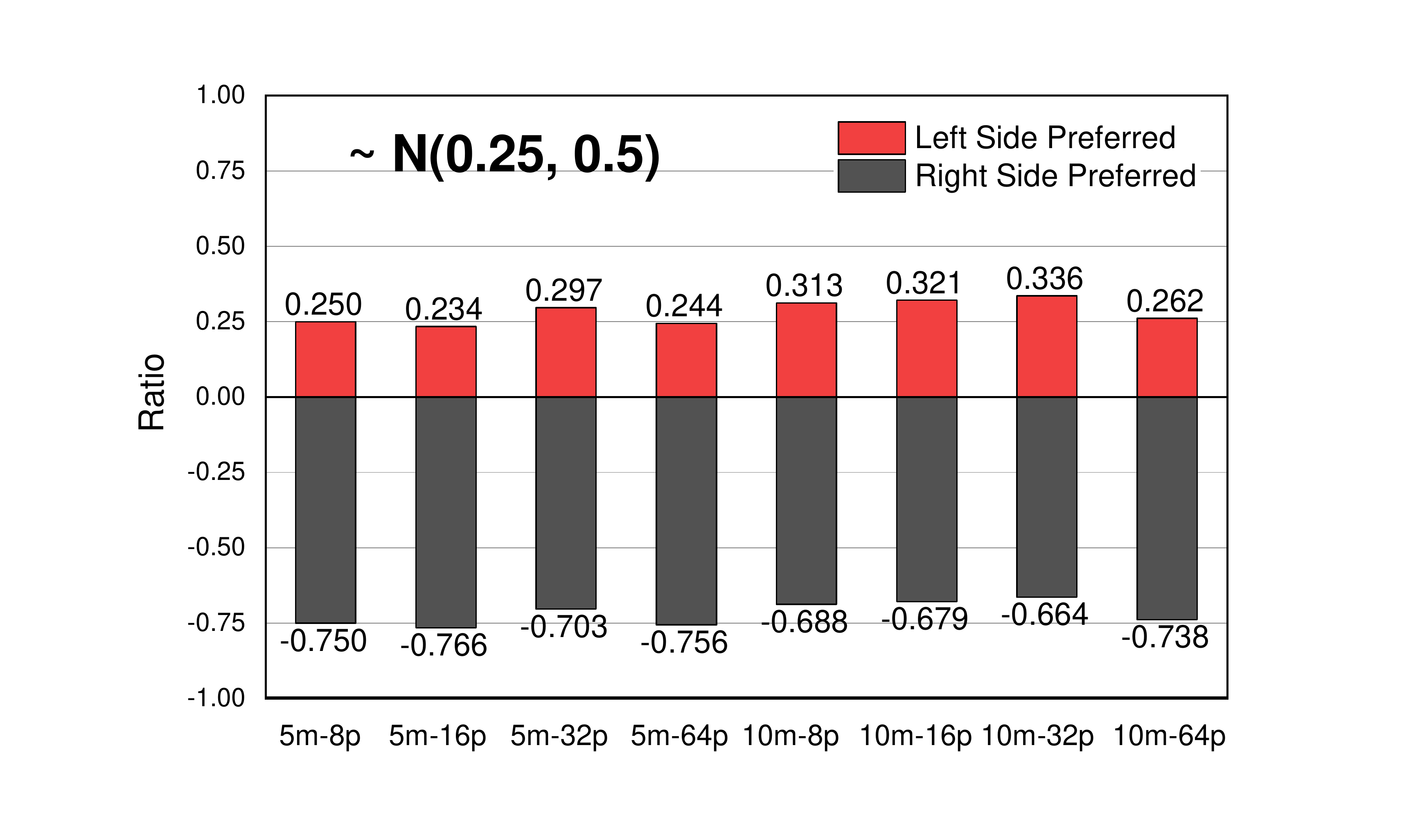}}\\
\subfloat[] { \includegraphics[width=0.9\textwidth]{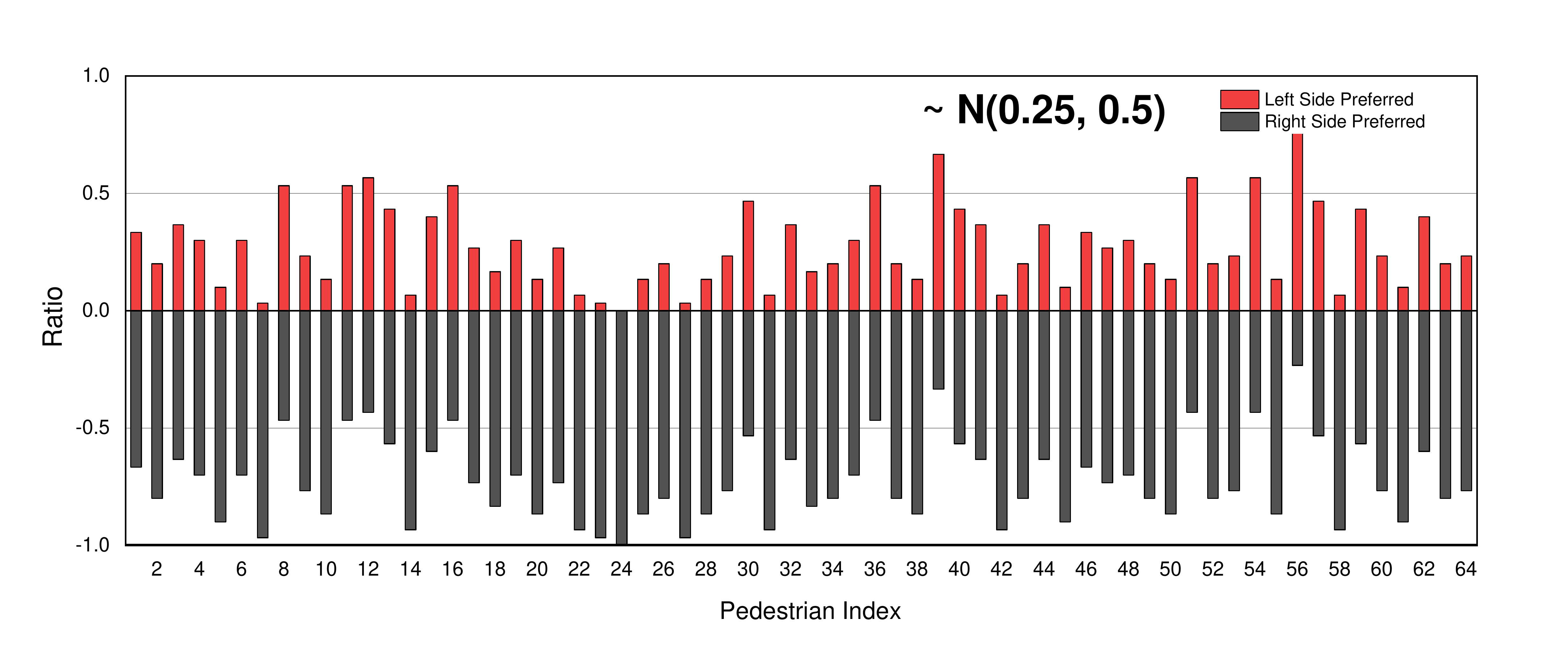}}
\caption{Simulation results in circle antipode experiments. (a) Right side preferred ratio distribution in simulation. (b) Right side preferred ratio distribution.} \label{figSideParmeters}
\end{figure}

\begin{figure}[!ht]
\centering{ \includegraphics[width=0.6\textwidth]{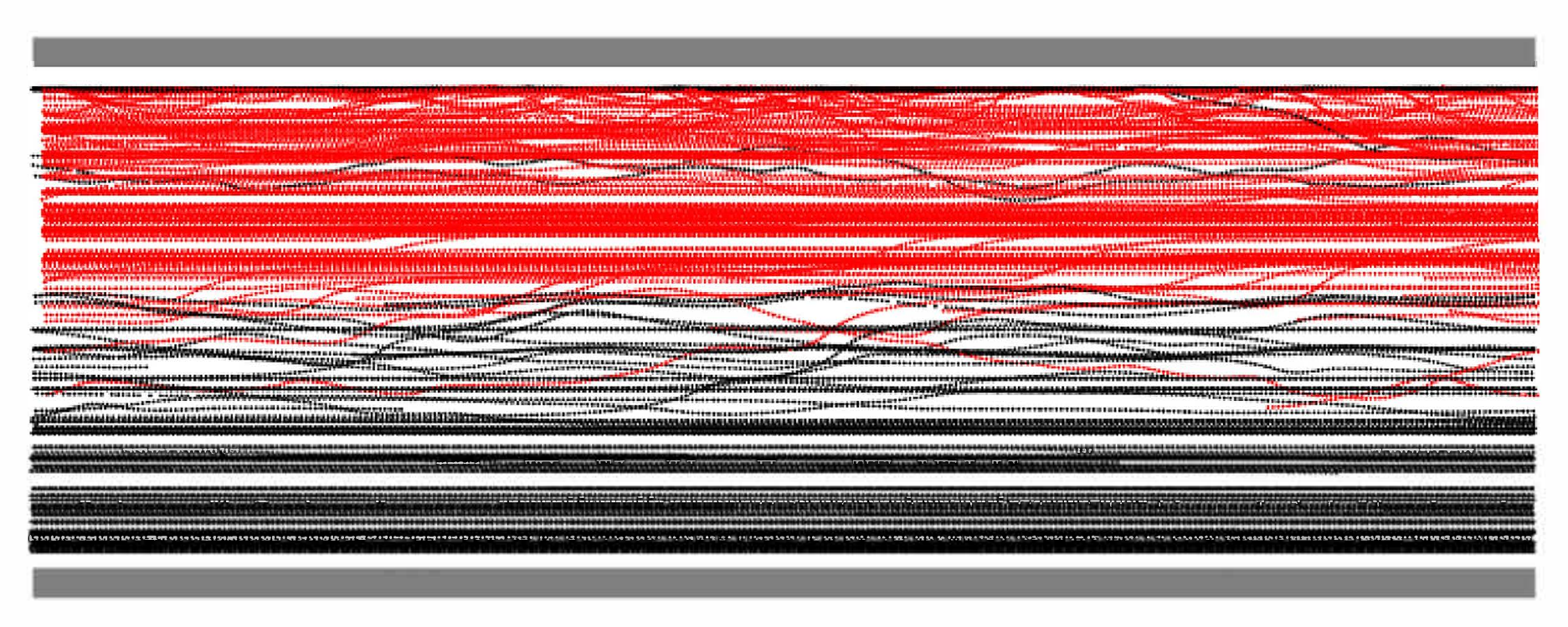}}
\caption{Simulated trajectories results in bi-directional flow corridor. The red points represent the pedestrian trajectories from right to the left, while the black points represent the pedestrian trajectories from left to right.} \label{figSideCorridor}
\end{figure}

First, the parameter $\gamma$ is attempted with a fixed value growing from -2.0 to 2.0, and the corresponding right side preferred probability increases from 0.2 to 0.8 as shown in Fig. \ref{figSidePreferenceProbabilitiy}. It also proves that the parameter $\gamma$ has a significant impact on the side preference phenomenon. Considering that the overall right side preferred probability is around 0.6875 in our circle antipode experiments, the mean value of $\gamma$ should be about 0.25 to 0.5. However, it is noted that pedestrians own heterogeneous tendencies for the side preference in reality (in Fig. \ref{figsideindividual}), hence distribution of $\gamma$ was proposed and attempted.

Here, the practical right side preferred ratios of the 64 participants in the series of experiments are cumulated in Fig. \ref{figsidepercentage}-a. Intuitively, the distribution of side preferred ratios own two peak values which indicate that different participants actually own different preferred side choices, and the right side preferred pedestrian numbers are obviously even greater. To quantitatively denote the differences between simulation results and experimental results, a variation index based on the right side preferred ratios is introduced as follows, 
\begin{equation}\label{eq difference}
Z = \sum_{i=1}^{n} \frac{(C_i - C_i^{exp})^2}{n}
\end{equation}
where $C_i$ and $C_i^{exp}$ denote the counts of participants with different right side preferred ratios in the simulations and the experiments, respectively. The normal distribution $N(\mu, \sigma)$ is introduced for the calibration of $\gamma$, and it is required to find the optimal parameters $\mu$ and $\sigma$ in the normal distribution. The calculated variation index $Z$ with two parameters $\mu$ and $\gamma$ can be found in Fig. \ref{figsidepercentage}-b, and the optimal parameters for the normal distribution are obtained as $\mu = 0.25$ and $\sigma = 0.5$.

By applying the optimal normal distribution of $\gamma$, the series of circle antipode experiments are simulated. The corresponding right side preferred ratio distribution is shown in Fig. \ref{figSideParmeters}-a, and a greater ratio of right side preferred individuals can still be found. Fig. \ref{figSideParmeters}-b presents the side preference of different individuals in the simulations. In all the 8 types of simulations, more pedestrians prefer to detour from the right side in the simulations, and the right side preferred pedestrians basically occupy around from 0.664 to 0.766, which are approximately the same with the experimental results. Also, the individual performance in the series of experiments are summarized in Fig. \ref{figSideParmeters}-c. It is found that 53 individuals in the simulations prefer to detour from the right side which also consistent with the individual results in empirical experiments.

What's more, other types of widely-used pedestrian movement situations, e.g., bi-directional pedestrian flow in corridor, are also attempted with the modified Voronoi model. In the bi-directional pedestrian flow experiment, 100 pedestrians are initialized in the $15 \times 5$ m corridor, and half of them head for the left side while the other half head for the right side. The specific distribution of $\gamma$ is applied and the simulation lasts for 5000 time steps. Fig. \ref{figSideCorridor} shows recorded trajectories of the corridor bi-directional flow in the last 1000 times steps. It's found that the pedestrians with different directions are basically seperated and most of the pedestrians are walking in the right side of the corridor, which is considered to be a well-known lane formation phenomenon.

\section{Conclusion and prospect}\label{section 5}

The side preference behaviors are investigated in the paper using the pedestrian trajectories from a series of circle antipode experiments. Benefits from the symmetric features of the experiments, the individual side preference results are calculated and analyzed. Our experimental results show that about 68.75\% of the pedestrians prefer the right side and the remaining 31.25 \% prefer the left side. Moreover, from a simple statistic analysis, it's believed that the side preference behavior is neither a simple deterministic behavior nor a stochastic process, and in fact, much more complicated factors are included in the behavior. A further statistical investigation on the related factors reveals that the handedness, gender, and height have no significant impact on the side preference behaviors, and more discussions about the side preference behavior mechanisms are still necessary.

Besides, it's found that the side choice is usually made at a very early stage through the research of the side preference consistency. The results imply that, in addition to the side preference behavior, the pedestrians actually make their choices about macroscopic motion strategies at the very beginning. It inspires us that a macroscopic route planning module seems to be necessary for the general pedestrian model to reproduce realistic crowd behaviors, especially in a complex environment. The statistics about the travel times of pedestrians with different side preferences demonstrate that the right side preferred pedestrians can get to their destinations faster, and the reason is clear that those right side preferred pedestrians are likely to meet fewer conflicts. It indicates that it is usually less conflicted and more efficient to choose a conformity motion pattern in crowds.

In the simulation part, a modified Voronoi model is formulated to carry out the circle antipode experiments, and a side preference parameter is introduced. To better agree with the experimental results especially the pedestrian heterogeneities, a normal distribution of the side preference is attempted and calibrated. The simulated side preference results of the circle antipode experiments agree with the experimental results, and further simulations of the bi-directional flow in the corridor also perform well.

A further problem is the quantitative influence of side preference behavior. It is interesting to see how the change of side preference ratio would affect the crowd efficiency in reality, and the circle antipode experiment can also be treated as a choice due to its conflicting situation and symmetry characteristics. 

\newpage

% \section*{Acknowledgment}\label{section acknowledgment}

% \begin{appendix}
% \setcounter{figure}{0}   %从零开始编号

% \section{Detailed simulations results} \label{section appendix a}

% % \begin{landscape}

% \newpage
% \end{appendix}

%\section*{References}

\bibliography{main}

\end{document}